\documentclass[twocolumn]{aastex63}
\usepackage{bm}

\newcommand{\mytilde}{\raise.19ex\hbox{$\scriptstyle\sim$}}

\usepackage[utf8]{inputenc}
\usepackage{CJK}
\usepackage{amsmath}
\usepackage{url}
\usepackage{lineno}
\usepackage{float}
\usepackage{placeins}
\usepackage{chngcntr}

\received{}
\revised{}
\accepted{}

\submitjournal{ApJ}

\shorttitle{Is Earendel a Star?}
\shortauthors{Scofield et al.}

\begin{document}
\title{Is Earendel a Star?: \\ Investigating the Sunrise Arc Using JWST Strong and Weak Gravitational Lensing Analyses}

\correspondingauthor{M. James Jee}
\email{zpscofield15@gmail.com, mkjee@yonsei.ac.kr}

\author[0009-0009-4086-7665]{Zachary P. Scofield}
\affiliation{Department of Astronomy, Yonsei University, 50 Yonsei-ro, Seoul 03722, Korea}
\author[0000-0002-5751-3697]{M. James Jee}
\affiliation{Department of Astronomy, Yonsei University, 50 Yonsei-ro, Seoul 03722, Korea}
\affiliation{Department of Physics and Astronomy, University of California, Davis, One Shields Avenue, Davis, CA 95616, USA}
\author[0000-0001-7148-6915]{Sangjun Cha}
\affiliation{Department of Astronomy, Yonsei University, 50 Yonsei-ro, Seoul 03722, Korea}
\author[0009-0007-7093-1758]{Hyosun Park}
\affiliation{Department of Astronomy, Yonsei University, 50 Yonsei-ro, Seoul 03722, Korea}
\begin{abstract}
     The galaxy cluster WHL J013719.8–08284 at $z = 0.566$ exhibits a strong-lensing feature known as the Sunrise Arc, which hosts the candidate star Earendel at $z\approx6.2$---the most distant star candidate observed to date. If this object is a star, or a system of a few stars, its apparent magnitude implies both extreme gravitational lensing magnification and unusually high luminosity. This study revisits Earendel's magnification, which, in previous literature, exhibits significant uncertainty across various lens models ($2\mu = 4{,}000$--$35{,000}$). We present an improved cluster mass reconstruction and a tighter constraint on Earendel's magnification using a joint strong- and weak-lensing analysis with JWST data. Our strong-lensing mass model, incorporating newly identified multiple-image systems from JWST imaging data and modifying the existing multiple-image assignment scheme, produces a root-mean-square (RMS) lens-plane scatter of less than $0\farcs3$. Additionally, our weak-lensing catalog achieves a source density of $\mytilde100$ galaxies arcmin$^{-2}$, providing constraints on the mass profile beyond the strong-lensing regime. In our best-fit model, we estimate the magnification of Earendel to be $\mu = 43$--$67$, significantly lower than previously proposed and thus calling into question its classification as a star.

\end{abstract}

\section{Introduction}\label{intro}

Gravitational lensing occurs when the gravitational potential of a massive object warps spacetime, altering the trajectories of light rays traversing therein. This lensing effect is observed as distortions in the images of astronomical objects positioned behind the lensing mass from the observer's perspective. There are several forms of gravitational lensing, with the two most relevant forms in analyses of galaxy clusters being weak-lensing (WL) and strong-lensing (SL). WL, the subtler of the two, involves slight shape distortions of background galaxies. It is a statistical measurement, utilizing overall distortion patterns across large ensembles of galaxies. SL is instead characterized by significant shape distortions, which occur when background sources align with the densest portions of the foreground mass. SL is therefore helpful for constraining the mass distribution in dense central regions of galaxy clusters and can also provide the necessary magnifying power to detect extremely distant, faint sources. 

An example of significant magnification from SL is MACS J1149 Lensed Star 1, nicknamed Icarus, observed in the field of galaxy cluster MACS J1149+2223. Given its redshift of $z=1.5$, \cite{kelly2018NatAs...2..334K} estimated that this star is magnified by a factor greater than 2,000---significant enough to resolve the single star with the Hubble Space Telescope. In this same observation, another compact object, the supernova SN Refsdal, is strongly lensed and observed as a multiple-image system. 

The subject of this study is the candidate star WHL0137-LS, nicknamed Earendel, which is hosted within the strongly lensed Sunrise Arc produced by the gravitational potential of the galaxy cluster WHL J013719.8–08284 (hereafter WHL0137-08). This candidate star is particularly interesting given its estimated redshift of $z=6.2 \pm 0.1$ \citep{salmon2020ApJ...889..189S}, which would make it the most distant isolated star observed to date. If it is a star, its apparent magnitude would require both extreme lensing magnification and exceptionally high luminosity.
 
 Earendel's classification as a star is based on a set of SL models produced in \cite{welch-2022Natur.603..815W} (hereafter W22a). These models predict that a lensing critical curve---a contour marking the regions of highest magnification in the lens plane---passes through or very near Earendel. As a result, its magnification is estimated to be $2\mu$ (the combined magnification of two unresolved counterimages) = $1{,}000$--$40{,}000$ (updated to $4{,}000$--$35{,}000$ in \cite{welchjwst2022ApJ...940L...1W}---hereafter W22b), with the significant uncertainty attributed to rapid changes in magnification near lensing critical curves. Reducing this uncertainty is vital to understanding the physical attributes of Earendel.
 
This large uncertainty in Earendel's magnification originally motivated this study, as incorporating WL data enables a more robust mass reconstruction of the cluster. Specifically, we leverage the more precise constraints of SL in the inner region of the cluster with the wider field coverage of WL to produce a more accurate mass reconstruction near the Sunrise Arc and a tighter constraint on the magnification of Earendel. To this end, we perform a combined WL+SL analysis for the WHL0137-08 cluster using a hybrid (combining parametric and free-form methods) lens modeling algorithm called \texttt{MrMARTIAN} (Cha et al. in prep). As mentioned in W22a, free-form components can probe a broader range of solutions for the mass model given the SL constraints. Importantly, such models can predict mass distributions that deviate from the typical mass-to-light assumption, which can provide substantial support for extreme magnification if they also yield results consistent with it.

A critical part of our analysis involves identifying and validating multiple-image systems in JWST imaging.  These systems play a direct role in reconstructing the mass distribution around the Sunrise Arc, thereby influencing the predicted magnification near Earendel. In addition to investigating previously proposed multiple-image systems, we introduce new candidate systems, which provide key insights into critical curve-crossing locations and, consequently, Earendel’s magnification.

 This paper is structured as follows: in Section \ref{sec2} we introduce the JWST dataset used in this work and discuss the relevant changes to the data reduction pipeline necessary for producing a clean image to be used for lensing analyses. In Section \ref{sec3} we introduce the gravitational lensing formalism. In Section \ref{sec4} we discuss aspects of the WL analysis in this work, including point spread function (PSF) modeling methods, shape measurement theory, and background source selection. Section \ref{sec5} provides details regarding the introduction of the SL dataset, with the results, discussion, and conclusion given in Sections \ref{sec6}, \ref{sec7}, and \ref{sec8}, respectively.

We assume a flat $\Lambda$CDM cosmology with $h=0.7$, $\Omega_{m} = 0.3$, and $\Omega_{\Lambda} = 0.7$. At the cluster redshift of $z=0.566$, the plate scale is $6.50 ~\rm kpc ~\rm arcsec^{-1}$. Masses are reported as $M_{500}$, which is the mass within a radius $R_{500}$ where the average density is 500 times the critical density of the universe at the redshift of the cluster. Unless otherwise stated, all right ascension and declination values in this work are referenced in the ICRS coordinate system.

\section{JWST Imaging}\label{sec2}
\begin{figure*}
    \centering
    \includegraphics[width=\textwidth]{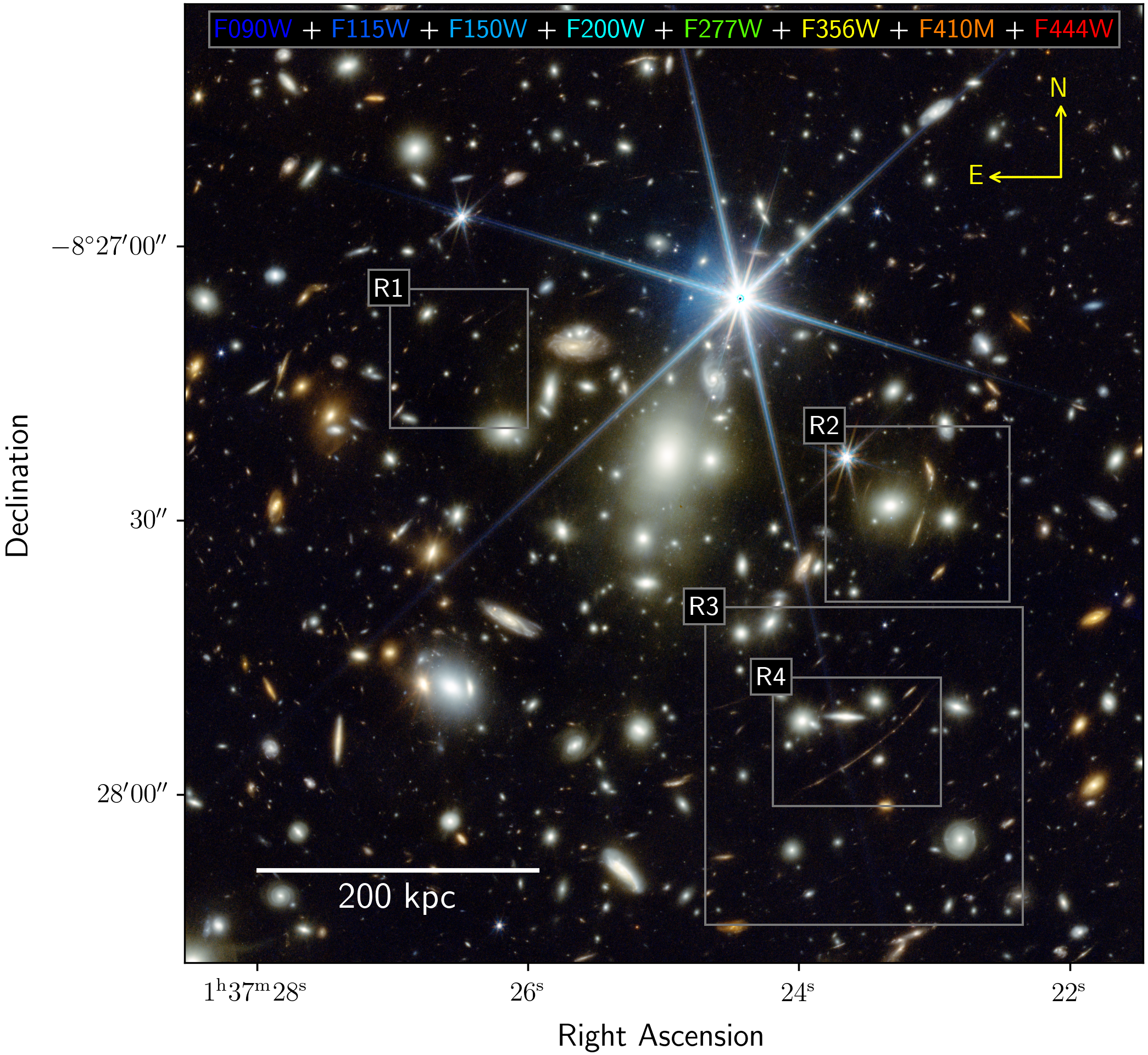}
    \caption{Eight-channel color-composite using all available NIRCam filters from the first epoch of the WHL0137-08 observation (GO 2282, PI Coe). The field of view shown lies within module B, which is centered on the cluster. The region labeled R1 encloses the multiple-image candidates shown in Figure \ref{extra_system}, while R2 and R3 correspond to the areas shown in the left and right panels of Figure \ref{halos}, respectively. R4 marks the region surrounding and including the Sunrise Arc, as shown in Figure \ref{sunrise_systems}.}
    \label{full_field}
\end{figure*}

WHL0137-08 was observed using the JWST NIRCam detector in July 2022 and January 2023 (GO 2282, PI Coe), with the first epoch of these observations being used in this work. This observation was carried out using eight filters (F090W, F115W, F150W, F200W, F277W, F356W, F410M, and F444W), with each filter having an integration time of 2104s \citep{bradley-2023ApJ...955...13B}. The NIRCam detector has two modules, module A and module B, separated by $40\farcs5$. The cluster is centered on module B, and the two modules have a combined coverage of $10.2 ~\rm arcmin^{2}$.
Among the available filters, we selected the F200W filter for our WL analysis for the reasons outlined in \cite{finnera-2023ApJ...953..102F}: this filter's PSF is critically sampled by the detector's native pixel scale of $\mytilde0\farcs031 ~\rm pixel^{-1}$, and it is the reddest of the short wavelength filters. Any filters redder than F200W use the long wavelength channel of the NIRCam detector, which has a native pixel scale of $\mytilde0\farcs063 ~\rm pixel^{-1}$, leading to a coarser PSF sampling. Figure \ref{full_field} provides a color-composite created using all eight filters, with notable regions used in subsequent figures labeled R1--R4.


The imaging data was reduced using the JWST Calibration Pipeline \citep{bushouse_2024_12692459} and coadded with a square kernel, \texttt{pixfrac} = 0.75, and an output pixel scale of $0\farcs02 ~\rm pixel^{-1}$ to mitigate the effects of undersampling \citep{finnerb-2023ApJ...958...33F}. Additional steps were implemented beyond the default pipeline, with one example being wisp removal---a process that utilizes templates provided in the JWST User Documentation \citep{2016jdox.rept......}. Wisps are extended, diffuse light features caused by stray light scattering within the telescope optics, often bypassing the primary mirror path and reaching the detector through unintended reflections. In weak-lensing analyses, it is necessary to remove these wisp features as they can interfere with the measurement of galaxy shapes---particularly for faint background sources.

Other steps added to the pipeline include exposure background subtraction and 1/\emph{f} noise correction, performed using modified versions of the \texttt{background\_subtraction} and \texttt{remstriping} algorithms provided by the CEERS team.\footnote{\url{https://github.com/ceers/ceers-nircam}}

Background subtraction is performed to ensure a final mosaic with a consistent background level. The 1/\emph{f} noise, characterized by higher power at lower spatial frequencies, is a more critical component to address. In the JWST NIRCam detector, this noise originates from the readout electronics and manifests as a striping pattern across images \citep{2016jdox.rept......}. As with wisp features, this striping must be mitigated to avoid contamination in the shape measurement process.

We have published our own modified pipeline, \texttt{young-jwstpipe},\footnote{\url{https://github.com/zpscofield/young-jwstpipe}} which streamlines the use of the default JWST calibration pipeline and incorporates the additional calibration steps needed to produce clean mosaic images. Further details regarding the data reduction process are provided in Appendix~\ref{app_data_reduction}.

\section{Gravitational Lensing Formalism}\label{sec3}
For a more detailed description of the gravitational lensing formalism, we refer the reader to \cite{Bartelmann:1999yn} and \cite{Schneider2006}, which serve as the main references for the theoretical framework discussed here.

To calculate lensing distortions, the lens equation should be solved for every point within a source, and is given as
\begin{equation}\label{lenseq}
    \bm{\beta} = \bm{\theta} - \frac{D_{\rm ds}}{D_{\rm s}} \bm{\hat{\alpha}} \left( D_{\rm d} \bm{\theta} \right) \equiv \bm{\theta} - \bm{\alpha}(\bm{\theta}) \, .
\end{equation}
Here, $\bm{\beta}$ is the angular position of a source in the source plane, $\bm{\theta}$ is the angular position of the image in the lens plane, and $\bm{\hat{\alpha}}$ is the physical deflection angle. The scaled deflection angle, $\bm{\alpha}$, is related to the physical deflection angle through the lensing geometry via the ratio $D_{\rm ds}/D_{\rm s}$ and is expressed in angular coordinates on the sky. The angular diameter distances $D_{\rm s}$, $D_{\rm d}$, and $D_{\rm ds}$ represent the distances from the observer to the source, from the observer to the lens (deflector), and from the lens to the source, respectively.

In Equation~\eqref{lenseq}, the deflection angle describes how a light ray's path is bent by a gravitational potential. In other words, given a source at angular position $\bm{\beta}$, one can observe the same source at any angular positions $\bm{\theta}$ that satisfy the lens equation. A key distinction between SL and WL is whether multiple images can be produced---that is, whether there are multiple solutions for $\bm{\theta}$ that satisfy the lens equation.

The dimensionless surface mass density $\kappa$, or convergence, is given by

\begin{equation}
    \kappa = \frac{\Sigma}{\Sigma_c} \, ,
\end{equation}
where $\Sigma_{(c)}$ is the (critical) surface mass density. $\Sigma_c$ is computed as follows:
\begin{equation}
    \Sigma_c = \frac{c^2 D_s}{4 \pi G D_d D_{ds}} \, .
\end{equation}
Here, $c$ is the speed of light and $G$ is the gravitational constant. The WL and SL regimes can be differentiated based on $\kappa$, as the SL regime is characterized by convergence $\kappa \gtrsim 1$.

The scaled deflection angle from \eqref{lenseq} can be written as
\begin{equation}\label{alpha}
    \bm{\alpha}(\bm{\theta}) = \frac{1}{\pi} \int_{\mathbb{R}^2} \mathrm{d}^2 \bm{\theta}' \kappa(\bm{\theta}') \frac{\bm{\theta} - \bm{\theta}'}{|\bm{\theta} - \bm{\theta}'|^2} \, ,
\end{equation}
which can be expressed as the gradient of the following deflection potential:
\begin{equation}\label{psi}
    \psi(\bm{\theta}) =  \frac{1}{\pi} \int_{\mathbb{R}^2} \mathrm{d}^2 \theta' \kappa(\bm{\theta}') \ln|\bm{\theta} - \bm{\theta}'| \, .
\end{equation}
This potential is a two-dimensional analogue of the Newtonian gravitational potential and thus satisfies the Poisson equation $\nabla^2 \psi(\bm{\theta}) = 2\kappa (\bm{\theta})$.

If a source is much smaller than the angular scale over which the lensing potential varies, the lens mapping can be linearized. In this regime, the image distortion is described by the Jacobian matrix:
\begin{align}\label{jacobian}
    A(\bm{\theta}) = \frac{\partial \bm{\beta}}{\partial \bm{\theta}} & = \left(\delta_{i,j} - \frac{\partial^2 \psi(\bm{\theta})}{\partial \theta_i \partial \theta_j} \right) \nonumber \\ & = \begin{bmatrix} 1-\kappa - \gamma_1&-\gamma_2\\-\gamma_2&1-\kappa + \gamma_1 \end{bmatrix} \, .
\end{align}
Here, $\delta_{i,j}$ is the Kronecker delta, which ensures that in the absence of lensing, $A(\bm{\theta})$ reduces to the identity matrix, yielding no distortion. The components $\gamma_1$ and $\gamma_2$ are the real and imaginary parts of the shear $\bm \gamma$, and relate to the deflection potential as follows:
\begin{equation}\label{gamma}
    \gamma_1 = \frac{1}{2}(\psi_{,11} - \psi_{,22}) \, , \quad \gamma_2 = \psi_{,12} \, ,
\end{equation}
where $\psi_{,ij}$ represents the second partial derivative of $\psi(\bm{\theta})$ with respect to $\theta_i$ and $\theta_j$.
Thus, we have the anisotropic shear $\bm \gamma$ and the isotropic convergence $\kappa$, which can describe the distortion of sources due to a gravitational potential. Using the relationship between $\psi$ and $\kappa$ in equation \eqref{psi} along with the definitions in equation \eqref{gamma}, we find that $\gamma$ can be related to $\kappa$ through
\begin{align}
    \bm {\gamma}(\bm{\theta}) & = \frac{1}{\pi}\int_{\mathbb{R}^2} \mathrm{d}^2 \bm{\theta}' \mathcal{D}(\bm{\theta}-\bm{\theta}') \kappa(\bm{\theta}') \, ; \nonumber \\
    \mathcal{D}(\bm{\theta}) & = \frac{-1}{(\theta_1 - i \theta_2)^2} \, .
\end{align}
Additionally, $\bm \gamma$ and $\kappa$ can be used to define the magnification $\mu$:
\begin{equation}\label{magnification}
    \mu = \frac{1}{\det(A)} = \frac{1}{(1-\kappa)^2 - |\bm \gamma|^2} \, ,
\end{equation}
with locations where $\det(A) = 0$ defining the critical curve.

In practice, we cannot measure $\gamma$ and $\kappa$ independently and instead observe the reduced shear $\bm{g}$, which can be expressed as a complex quantity
\begin{equation}\label{g}
    \bm{g} = g_1 + i g_2 = \frac{\bm \gamma}{(1-\kappa)} \, ,
\end{equation}
where $g_{1,2}$ are the first and second components of the reduced shear, respectively. With this definition, \eqref{jacobian} becomes
\begin{equation}\label{g_jac}
    A(\bm{\theta}) = (1-\kappa)\begin{bmatrix} 1 - g_1 & -g_2 \\ -g_2 & 1 + g_1 \end{bmatrix}
 \, .
\end{equation}
In the SL regime, where distortions are highly nonlinear and $|\bm g|>1$, we replace $\bm g$ with $1/\bm{g}^*$, where $\bm{g}^*$ is the complex conjugate of $\bm g$.

\section{Weak lensing analysis}\label{sec4}

\subsection{PSF Modeling}

Given the small distortions involved in WL analyses, accounting for image distortions caused by the telescope is essential. The PSF captures these intrinsic, lens-independent distortions as the response of the telescope to a point source of light. For the JWST, there are various methods for modeling the PSF, with the choice for this work being a principal component analysis (PCA). This method defines a set of basis functions derived from the stars observed in the field, enabling the reconstruction of the PSF as represented by these point sources. PCA has been demonstrated to be effective in numerous previous lensing studies \citep[e.g.,][]{2007PASP..119.1403J, finnera-2023ApJ...953..102F, Cha_2024,hyeonghan2024arXiv240500115H, ahn2024substructuressubstructurescomplexpostmerging}. Another method involves the \texttt{STPSF}\footnote{\url{https://stpsf.readthedocs.io/en/latest/}} simulation package \citep{webbpsf} and uses optical path difference (OPD) maps measured onboard the JWST near the date of observation to simulate PSFs on any detector. These OPD maps describe how light paths are altered relative to the ideal optical state as prescribed in the Fraunhofer approximation, effectively encoding the various sources of distortion present. This simulation method was utilized for a recent analysis in \cite{Cha_2024}.

\begin{figure*}[!t]
\centering
	\includegraphics[width=\textwidth]{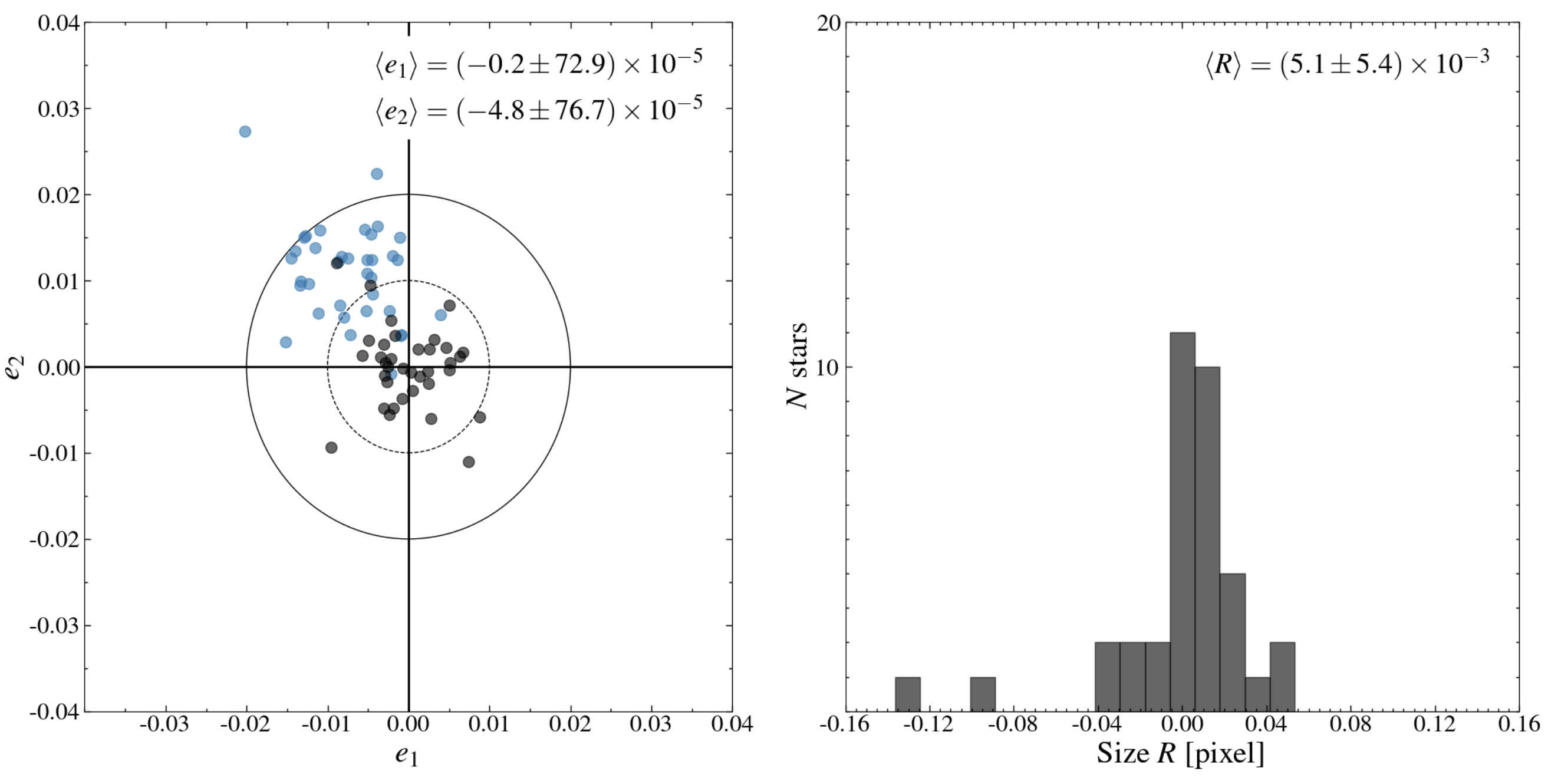}
	\caption{PSF model corrections for ellipticity and size. Left: Complex ellipticity components of stars measured in the WHL0137-08 field (blue) and residual ellipticities (black) computed as the difference between observed and model values. The inner dotted circle represents $|e|=0.01$, while the outer solid circle corresponds to $|e| = 0.02$. Right: Residual size $R$ between observed stars and the corresponding reconstructed stars.}
	\label{psf_ell}
\end{figure*}

The PCA model is derived directly from the data, making it effective in capturing the true observed distortion in a given observation. However, this method relies on having a sufficient number of stars with a relatively uniform distribution across the field to produce an accurate model. While the \texttt{STPSF} model does not have this limitation, the simulation may produce a model showing different features than those of the observed PSF. \cite{Cha_2024} demonstrated that given an adequate number and distribution of stars, both modeling methods perform similarly. In that work, we used a custom implementation of \texttt{STPSF} called \texttt{stpsf-mosaic}\footnote{\url{https://github.com/zpscofield/stpsf-mosaic}}, which we developed to streamline the production of accurate \texttt{STPSF} simulation PSF models for mosaic images. This implementation incorporates the exposure times and orientations of all contributing exposures to generate a PSF at each specified position within the observation field.

We select 36 high-quality stars with a largely uniform distribution from the WHL0137-08 field and employ the empirical PCA modeling method for this work. To create the PSF model, we compute the mean star from those selected in the field and subtract it from each $31\times31 ~\rm pixel$ star stamp to generate a $31 \times 31 \times 36$ $(n_{\mathrm{stars}})$ residual array. We perform a PCA on the flattened $36\times961~\rm pixel$ residual array, transforming the residual array to an orthogonal basis, and retain the first 16 principal components which capture $\mytilde90\%$ of the variance in the mean PSF. This choice enables the model to reproduce the ellipticity of the observed PSFs while minimizing the influence of noise.


To measure the effectiveness of our PSF model, residual shape parameters are computed between observed stars in the field and PSFs reconstructed at these star locations using the model. These shape parameters are measured using the following quadrupole moments \citep{2007PASP..119.1403J}:
\begin{align}
    Q_{i,j} = \frac{\int \mathrm{d}^2 \bm{\theta} W(\bm{\theta})I(\bm{\theta})(\theta_i - \bar{\theta}_i)(\theta_j - \bar{\theta}_j)}{\int \mathrm{d}^2 \bm{\theta} W(\bm{\theta})I(\bm{\theta})} \;, \nonumber \\ \quad i,j \in {1,2} \;.
\end{align}
Here, $I(\bm{\theta})$ is the pixel intensity at $\bm{\theta}$, $W(\bm{\theta})$ is a circular Gaussian weight function utilized to suppress noise in the outer regions of the PSF, and $\bar{\theta}_{i,j}$ is the center of the star. This center is defined as
\begin{equation}
    \bm{\bar{\theta}} = \frac{\int \mathrm{d}^2 \bm{\theta} q_I [I(\bm{\theta})] \bm{\theta}}{\int \mathrm{d}^2 \bm{\theta} q_I [I(\bm{\theta})]} \, ,
\end{equation}
with $q_I[I(\bm{\theta})]$ being a weight function defined as $q_I[I(\bm{\theta})] = I(\bm\theta)$ in this case. This choice of weight function results in $\bar{\theta}$ being the center of light \citep{Bartelmann:1999yn}.
Using the quadrupole moments, we can define the complex ellipticity components ($e_1$, $e_2$) and size ($R$):
\begin{align}
    e_1 + i e_2 & = \frac{Q_{11} - Q_{22} + 2iQ_{12}}{Q_{11}+Q_{22}+2(Q_{11}Q_{22}-Q^2_{12})^{\frac{1}{2}}} \;, \\ 
    R & = \sqrt{Q_{11} + Q_{22}} \;.
\end{align}
For an ellipse with semi-major and -minor axes $a$ and $b$ and position angle $\phi$ measured counterclockwise from the first image axis, $e_1$ and $e_2$ correspond to
\begin{equation} \label{ellipticity}
    e_1 = \frac{a-b}{a+b} \cos(2\phi) \;, \quad e_2 = \frac{a-b}{a+b} \sin(2\phi) \, ,
\end{equation}
with $(a-b)/(a+b)$ referred to as the ellipticity $e$.

We compare the residual ellipticities and sizes between the observed stars and the stars reconstructed using the PSF model in Figure \ref{psf_ell}. The left panel presents a target plot with residual ellipticities (black points) closely centered around (0,0), with mean and standard error values $\langle e_1 \rangle = (-0.2 \pm 72.9) \times 10^{-5}$ and $\langle e_2 \rangle = (-4.8 \pm 76.7) \times 10^{-5}$. These small residuals demonstrate that the PSF model can effectively capture the ellipticity of the observed stars. The residual sizes have a median and standard error of $\langle R \rangle = (5.1 \pm 5.4) \times 10^{-3}$, similarly indicating that the PSF model can reconstruct the characteristic size of observed stars in the field.

\subsection{Shape Measurement}
We measure the shape of each selected background galaxy in the image by using a forward-modeling technique, which involves fitting an elliptical Gaussian profile to the observed brightness distribution. Starting with shape estimates from Pythonic Source-Extractor \citep{Barbary2016,bertin-1996A&AS..117..393B}, we generate an elliptical Gaussian function with rotation angle, amplitude, and semi-minor and -major axes lengths as free parameters while fixing the background and centroid values. The Gaussian function is then convolved with an appropriate PSF and fit to the surface brightness distribution of the observed galaxy with the \texttt{MPFIT} minimization \citep{mpfit}. Noise is accounted for by creating a root-mean-square error map from the weight map produced by the image calibration pipeline. The ellipticity and position angle for each background galaxy are defined in the same way as in equation \eqref{ellipticity}.

This elliptical Gaussian fitting algorithm is subject to bias, as discussed in \cite{finnera-2023ApJ...953..102F}. In our analysis, we simply adopt the multiplicative bias from this same work (1.11 and 1.07 for $e_1$ and $e_2$, respectively) produced with the SFIT shear calibration technique (HyeongHan et al. in prep).

\subsection{Source Selection}
Gravitational lensing distorts only the images of objects behind the lens, making a clean sample of background sources essential. Many strongly lensed galaxies are immediately apparent, though additional multiple-image systems can often be identified through color analyses, redshift estimates, or lens modeling. In contrast, identifying weakly lensed sources is more challenging, as the small distortion effects are indistinguishable for individual galaxies. These subtle distortions are instead measured statistically by averaging over large ensembles of galaxies, overcoming intrinsic shape noise. In our work we used Source-Extractor to create a source catalog, with photometric redshifts employed for categorizing galaxies as background sources. The photometric redshifts for galaxies in the WHL0137-08 cluster field were accessed through the Cosmic Spring public repository\footnote{\url{https://cosmic-spring.github.io/index.html}}, with a description of the photometric redshift estimation procedure given in \cite{bradley-2023ApJ...955...13B}. 

To classify as a background source, we constrain the $1\sigma$ lower limit of a galaxy's redshift probability distribution to above $z=0.626$. This value is six times the typical line-of-sight velocity dispersion of a massive merging cluster \citep{Golovich_2019,finnera-2023ApJ...953..102F} beyond the cluster at $z=0.566$. To ensure that we used galaxies with reliable redshift estimates, we required each galaxy to have a best-fit redshift of $z<7.0$ and an uncertainty of $\delta z < 25\%$. Additionally, during shape measurement, we exclude sources with ellipticity uncertainty $\delta e \geq 0.3$ and ellipticity $e>0.85$, with the latter often signifying a spurious detection.
The final source count for the total field is 1183 across both modules, with 677 in module A and 506 in module B, corresponding to source densities of $\mytilde132.7 ~\rm arcmin^{-2}$ and $\mytilde99.2 ~\rm arcmin^{-2}$, respectively. The final source density is lower than the $\mytilde350 ~\rm arcmin^{-2}$ density reported for the Abell 2744 WL analysis by \cite{Cha_2024} due in part to differences in the F200W limiting magnitude. For the WHL0137-08 observation, the F200W 5$\sigma$ limiting AB magnitude is 28.7 in an $r=0\farcs1$ circular aperture \citep{bradley-2023ApJ...955...13B}, resulting from an exposure time of 2,104 seconds. In comparison, the Abell 2744 F200W mosaic---with a significantly longer exposure time of 3.7 hours---reaches a deeper limit of 30.12 AB mag in an $r=0\farcs08$ circular aperture \citep{bezanson2024ApJ...974...92B}.

\section{Strong Lensing Analysis}\label{sec5}  

\begin{figure*}[tp]
\centering
	\includegraphics[width=\textwidth]{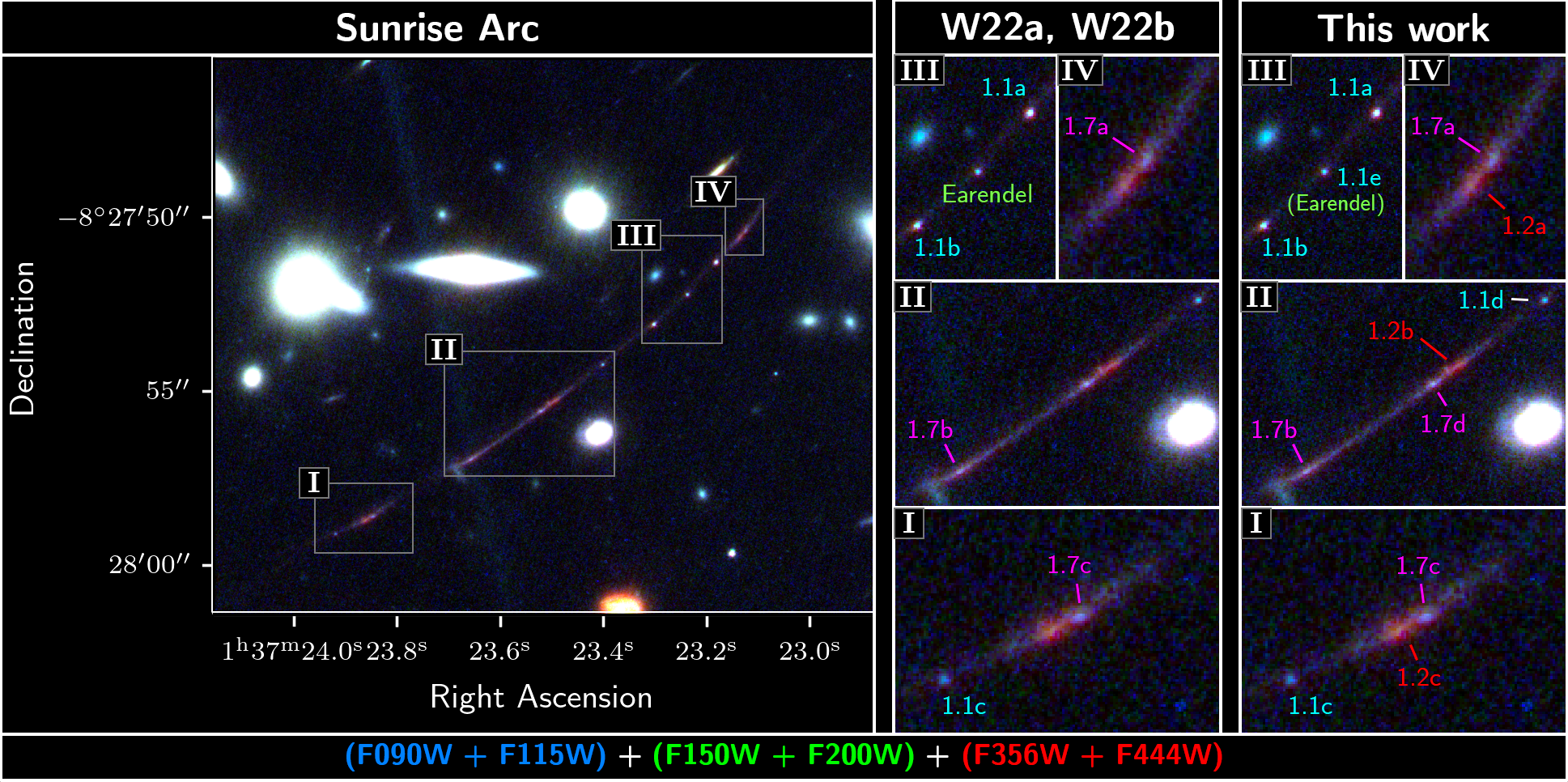}
	\caption{Region R4 from Figure \ref{full_field}, showing multiple-image systems within the Sunrise Arc according to W22a, W22b, and this work. Systems 1.2 and 1.7 show consistent colors between clumps, following the expected parity. We assign the same label to the multiple images within system 1.1 for consistency with W22a and W22b. However, we explore the possibility that they may not belong to the same system due to differences in their colors. For example, image 1.1c has a different color compared to 1.1a and 1.1b, and its color more closely resembles that of 1.1d and 1.1e (Earendel).}
	\label{sunrise_systems}
\end{figure*}

\subsection{Multiple-image systems}\label{mult_syst}
Candidate multiple images can be identified through visual inspection, photometric redshift estimations, or lens model predictions, with the assignment of multiple-image systems directly influencing the resulting mass reconstruction. This process is especially important for the sparse WHL0137-08 SL dataset, as it significantly affects the inferred mass distribution near the Sunrise Arc. W22a assigned three multiple-image systems in their analysis, two of which are part of the Sunrise Arc and are shown in the middle panel of Figure \ref{sunrise_systems}. These systems were identified using HST imaging data, and while W22b utilized JWST data to improve constraints on Earendel's magnification and maximum radius, no changes were made to their multiple-image system assignments or lens models. 

\begin{figure}[tp]
    \vspace{3mm}
    \centering
    \includegraphics[width=0.95\columnwidth]{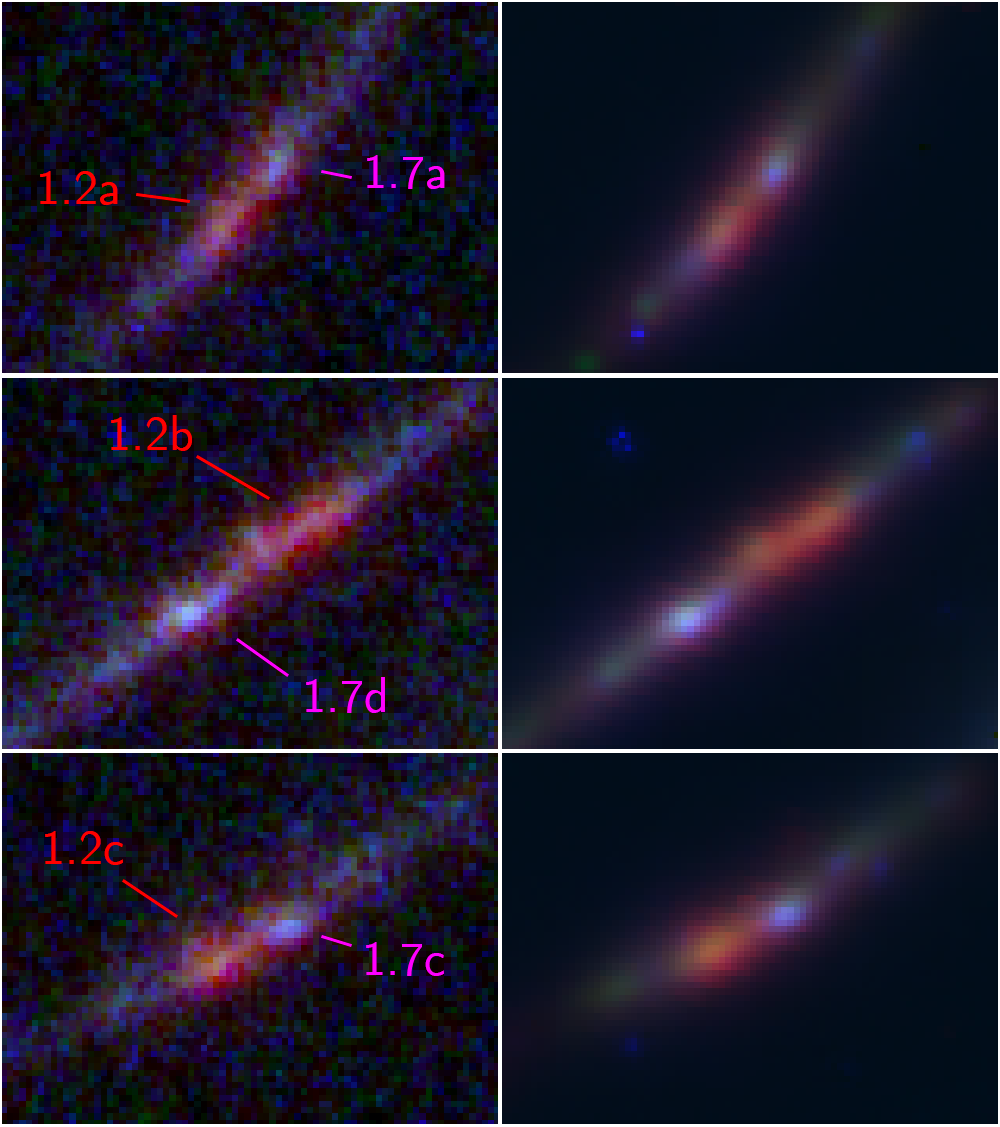}
    \caption{Zoomed-in views of systems 1.2 and 1.7(a, c, d). Both the morphology and color suggest that the three 1.1–1.7 pairs belong to the same system. This identification is further aided by the denoised version (right), although the feature is still visible in the original color image (left).}
    \label{denoised_sunrise}
\end{figure}

Utilizing the JWST imaging data, we have identified new multiple-image candidates within the Sunrise Arc that are likely to be true multiple images based not only on a visual inspection but also on the resulting lens-plane scatter found during modeling (discussed in Sections \ref{model_criteria} and \ref{wlslresults}). Figure \ref{sunrise_systems} shows the multiple-image systems within the Sunrise arc from W22a and W22b, as well as our own proposed multiple-image systems. When assigning F090W+F115W, F150W+F200W, and F356W+F444W to the blue, green, and red channels, respectively, we see distinct clumps with very different colors, which exhibit the expected parity along the arc (1.7a+1.2a; 1.7d+1.2b; 1.7c+1.2c). To investigate these clumps further, we denoised the WHL0137-08 field in the region surrounding the Sunrise Arc, with the results (utilizing the same filter combination as shown in Figure \ref{sunrise_systems}) for multiple-image systems 1.7 and 1.2 shown in Figure \ref{denoised_sunrise}. While this denoising process highlights the morphological similarities between the three clumps, their distinct substructures are clearly visible in the original color image as well. The denoising procedure utilizes a Restormer \citep{zamir2022restormer} deep learning architecture as described in \cite{Park_2024} and is discussed in detail in Appendix \ref{denoising_desc}.

In addition to these newly identified clumps and sub-clumps, we find that image 1.1d has a more similar color to 1.1c as opposed to 1.1a or 1.1b (with 1.1a, 1.1b, and 1.1c forming a multiple-image system in W22a). Additionally, given the observed parity along the arc, we propose that image 1.1e (Earendel) may be a part of the same system as 1.1c and 1.1d, based on its position and color. Importantly, these multiple-image systems differ from those used in W22a and can significantly change the predicted location of critical curve crossings along the Sunrise Arc.

We also identify new multiple-image system candidates, shown in Figure \ref{extra_system}, in the northwest region of the field. Images 3a and 3b both have best-fit photometric redshifts of $z\approx6.6$ according to estimates from the Cosmic Spring team \citep{bradley-2023ApJ...955...13B}, and they also exhibit very similar colors. The redshift for image 3a is derived from its second-brightest clump, as no valid fit was obtained for the brightest clump. The multiple images in system 4 lie at a similar radial distance from the cluster center as those in system 3, and the two systems are tangentially aligned, suggesting they may belong to different sections of the same galaxy. Moreover, the images in both systems exhibit opposite parity along their shared tangential axis. Moreover, images within both systems exhibit opposite parity along their shared tangential axis. However, we do not include this system in our SL analysis given its distance from the Sunrise Arc and unconfirmed status as a true multiple-image system, and instead identify it for potential use in future studies. 

We list our multiple-image system candidates in Table \ref{multiple_images} and summarize our proposed changes to the multiple-image system assignment within the Sunrise Arc as follows:
\begin{itemize}
    \item We identify a new image (1.7d), which is likely a part of the previously identified system 1.7.
    \item We identify system 1.2, composed of red sub-clumps that exhibit clear parity along the arc in combination with system 1.7.
    \item Based on this same parity, we propose that image 1.1d is more likely to be a counterpart to image 1.1c as opposed to images 1.1a or 1.1b. This proposition is further supported by the relative colors between images within system 1.1, with 1.1c exhibiting a more similar color to 1.1d than 1.1a or 1.1b.
    \item We propose that image 1.1e (Earendel) may be a part of the same system as images 1.1c and 1.1d, though we are less certain of this association than we are about 1.1d being a multiple image of 1.1c.
\end{itemize}

\subsection{Lens modeling algorithms}
Lens modeling is often performed using either parametric or free-form algorithms. Parametric lens modeling involves placing analytic profiles at the locations of cluster member galaxies and using SL multiple-image positions to constrain the model parameters. This approach has the advantage of requiring significantly fewer free parameters than free-form methods, reducing computational complexity and diminishing the risk of producing unphysical mass distributions. However, due to their inherent lack of flexibility, purely parametric models suffer from underfitting and may not capture the true complexity of cluster mass distributions. In contrast, free-form algorithms have no specific functional form for the mass distribution and treat each grid cell of the convergence map as a free parameter, with each value of $\kappa$ determined through a minimization procedure. While a large number of free parameters can result in longer execution times and potential overfitting resulting from limited SL constraints, the flexibility of free-form methods can accommodate density fluctuations in mass distributions that parametric models might miss.

WHL0137-08, in particular, poses a significant challenge for SL mass reconstruction due to its limited multiple-image systems. While parametric methods can still produce usable models, albeit with a risk of oversimplification or bias, free-form methods suffer significantly more as they require a substantial number of well-distributed multiple-image systems to prevent overfitting and ensure physically plausible mass reconstructions.

To address this lack of SL constraints, we utilize the new hybrid parametric and free-form mass reconstruction algorithm \texttt{MrMARTIAN} (Multi-resolution MAximum-entropy Reconstruction Technique Integrating Analytic Node, Cha et al. in prep) to combine WL and SL datasets. This code is adapted from \texttt{MARS} \citep{Cha_2022, cha_2023ApJ...951..140C}, which is a free-form lens modeling algorithm proven effective in previous WL and WL+SL analyses \citep{cha_2023ApJ...951..140C, Cha_2024, hyeonghan_2024ApJ...962..100H, pascale_2024arXiv240318902P}. The \texttt{MrMARTIAN} algorithm incorporates a parametric component, allowing truncated pseudo-elliptical Navarro-Frenk-White (TNFW) profiles to be positioned in the reconstruction field. This approach is particularly useful when SL multiple-image systems alone cannot properly reconstruct the mass distribution in high-convergence regions. By combining the stability of a parametric method with the flexibility of a free-form approach, \texttt{MrMARTIAN} can accommodate the complex nature of lensing mass distributions while also preventing the overfitting and unphysical mass distributions occasionally seen in free-form methods. 

The free-form component utilizes the WL dataset to aid in the reconstruction of the large-scale mass distribution of WHL0137-08. Additionally, the SL dataset informs the free-form component in high-density regions of the cluster, where there may be mass fluctuations and substructures that analytic profiles cannot adequately represent. The parametric component incorporates constraints from both datasets and anchors the lens model to a more physically plausible configuration, while also accounting for the higher mass densities in the SL regime that the free-form component cannot model effectively due to the limited number of SL constraints. The complimentary nature of not only the WL and SL datasets but also the free-form and parametric components allows us to overcome the limitations posed by the sparse SL dataset in the WHL0137-08 field.

\begin{figure*}[tp]
\centering
    \includegraphics[width=\textwidth]{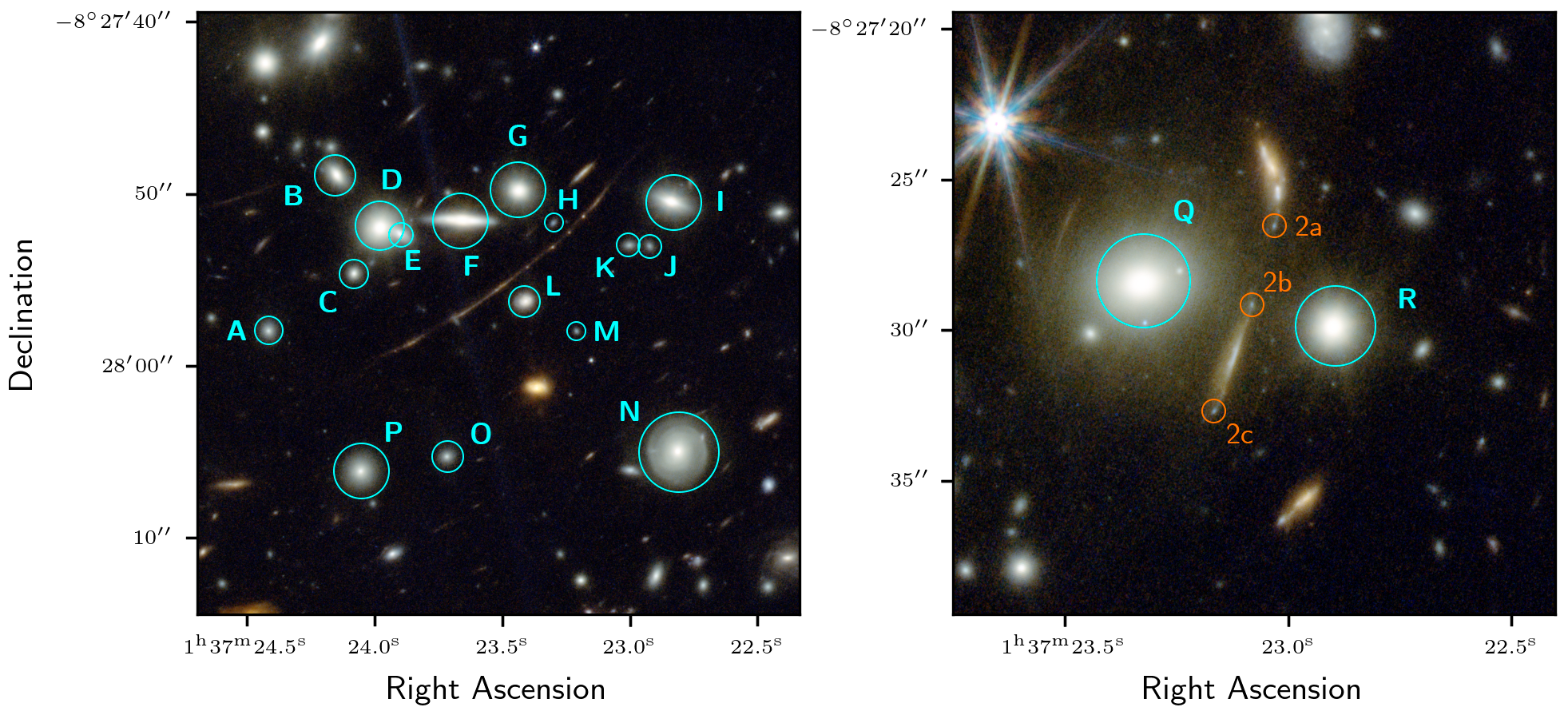}
    \caption{Regions R2 and R3 from Figure \ref{full_field}. Left: galaxies near the Sunrise Arc used in lens modeling, highlighted in cyan. All galaxies were selected as cluster members based on photometric redshifts from \cite{bradley-2023ApJ...955...13B}, except galaxies H and L, which are included due to their proximity to the arc. Right: additional cluster members Q and R, highlighted in cyan, used in modeling only when considering mutliple-image system $2$, enclosed in orange.}
    \label{halos}
\end{figure*}

\subsection{Model selection criteria}\label{model_criteria}

SL modeling aims to converge multiple-image systems to a single location in the source-plane while properly reproducing the observed multiple-images in the lens-plane. Therefore, to assess the accuracy of our models, we utilized two key metrics: multiple-image source-plane scatter ($\Delta_{\rm rms, \, source}$) and lens-plane scatter ($\Delta_{\rm rms, \, lens}$). $\Delta_{\rm rms, \, source}$ quantifies the rms deviation between the predicted source-plane positions and their mean, as the true source position is unknown, while $\Delta_{\rm rms, \, lens}$ provides the rms deviation between the observed and predicted positions of multiple-images in the lens-plane. The latter metric is described in \cite{Cha_2024}:

\begin{equation}\label{eqn_rms}
    \Delta_{\rm rms, \,lens}=\sqrt{\frac{1}{M}\sum_{m=1}^{M}|\bm{\theta}_{\mathrm{truth},m}-\bm{\theta}_{\mathrm{model},m}|^{2}} \, ,
\end{equation}
where $\bm{\theta}_{\mathrm{truth},m}$ represents the observed image positions. The source-plane scatter follows the same form but replaces $\bm{\theta}_{\mathrm{truth},m}$ with the average predicted source position $\langle \bm{\theta}_{\mathrm{model},m} \rangle$. 

An accurate model should exhibit low scatter in both the source- and lens-plane, and satisfying one condition does not necessarily imply the other. The source-plane scatter, for instance, can be artificially minimized by increasing lensing magnification, potentially introducing a magnification bias. Specifically, higher magnifications reduce source-plane scatter without necessarily improving the physical accuracy of the model. Moreover, a reduction in source-plane scatter does not necessarily correlate with a reduction in lens-plane scatter due to the nontrivial nature of lensing transformations. 

The relationship between small positional perturbations in the source-plane and their corresponding displacements in the lens-plane is governed by the Jacobian matrix \eqref{jacobian}, as described by
\begin{equation}\label{inv_jac}
    \partial \bm{\theta} = A^{-1}(\bm{\theta}) \partial \bm{\beta} \, .
\end{equation}
The lensing magnification, given in Equation \eqref{magnification}, is related to this transformation due to its dependence on the determinant of the Jacobian. As magnification increases, this determinant approaches zero, making $A^{-1}(\bm{\theta})$ ill-defined. This leads to an increasingly unstable lens mapping \eqref{inv_jac}, where small deviations in source-plane positions can result in significant displacements in the lens plane. By considering both source- and lens-plane scatter, we can select models that produce expected outcomes in both planes while mitigating magnification bias due to overfitting.

\subsection{TNFW profile selection}\label{tnfw_halo_sec}

When adding TNFW profiles to the reconstruction field, we only used galaxy positions near the multiple images within the Sunrise Arc. We justify this choice as follows:
\begin{itemize}
    \item While TNFW profiles could theoretically be added at every cluster-member position, the limited number of multiple-image systems would be unable to properly inform the parameters of each profile. These parameters use flat priors unrelated to the observed characteristics of their corresponding galaxies.
    \item Given the truncation of TNFW profiles, those positioned far from the Sunrise Arc yield a minimal contribution in the region surrounding the arc and, consequently, have a negligible effect on the magnification estimate of Earendel. Moreover, any diffuse contribution from distant galaxies should be sufficiently captured by the grid component of \texttt{MrMARTIAN}. 
    \item Adding too many TNFW profiles in the reconstruction field can lead to degeneracies between the grid and parametric components of \texttt{MrMARTIAN}. The TNFW profiles are primarily intended to aid in reconstructing the cluster mass in regions with SL features, particularly where there are insufficient constraints for a fully free-form modeling approach to be effective.
\end{itemize}
Although system 2 lies far from the Sunrise Arc, W22a noted its importance in reproducing the arc's extent in their models. To enable fair comparisons, we also produced models including this system and nearby cluster-member galaxies. The galaxy positions used for modeling are shown in Figure \ref{halos}. All TNFW profiles used in modeling are centered on these galaxies, which are member galaxies selected based on photometric redshifts, except for galaxies H and L. These were included in some models due to their proximity to the Sunrise Arc and their role in the W22a analysis.

\begin{figure*}[tp]
\centering
	\includegraphics[width=\textwidth]{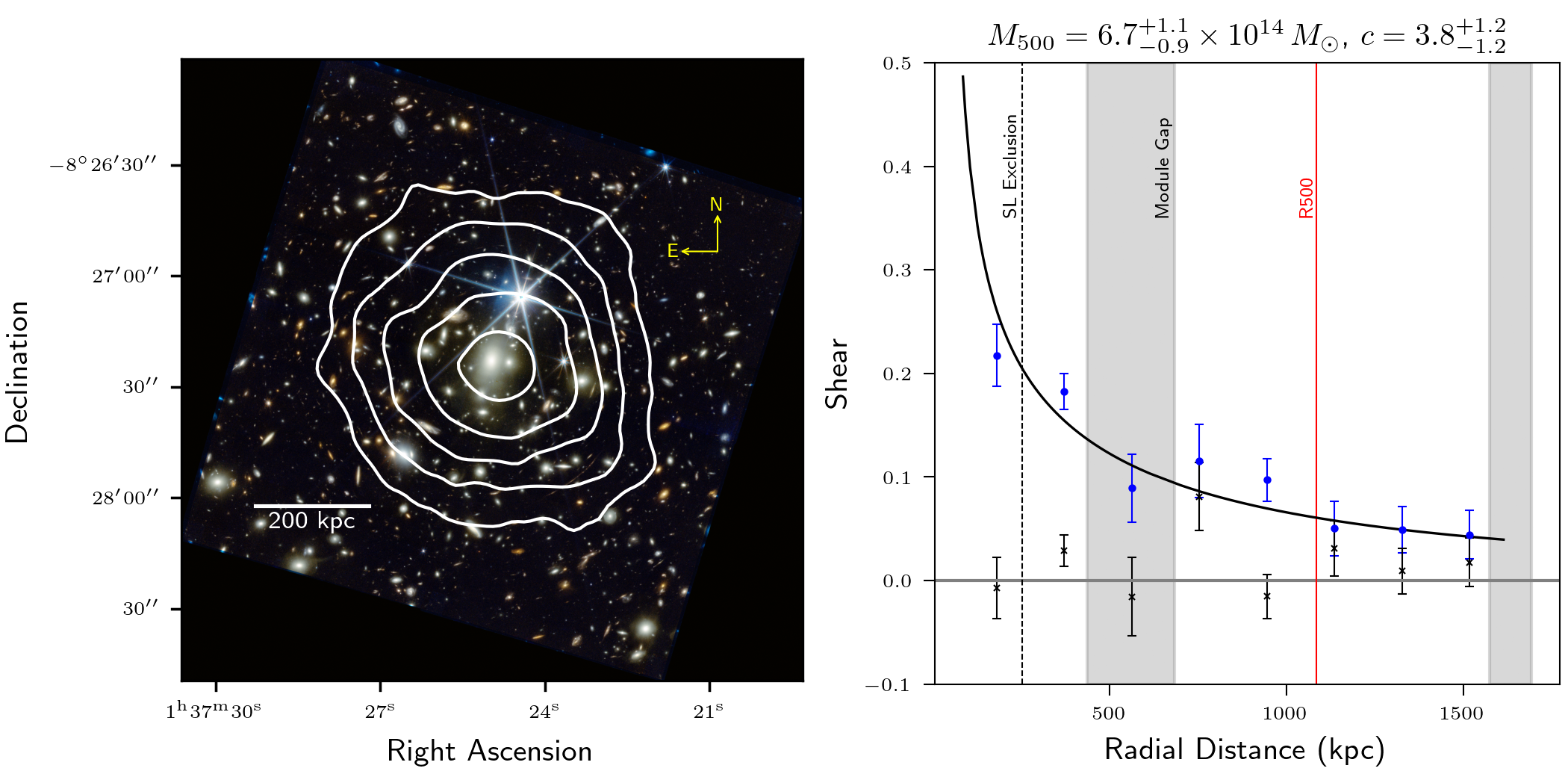}
	\caption{Left: Convergence map contours overlayed on the color-composite image of module B. The contours start at $3\sigma$ and increase in $1\sigma$ steps. The convergence map is smoothed with a Gaussian kernel where $\sigma \approx 2\farcs5$. Right: Binned tangential shear (blue dots) and cross shear (black crosses) plotted against radial distance from the BCG. The dotted line represents the radius from the BCG excluded in the fitting---as to avoid the SL regime---while the red line is at $R_{\mathrm{500}}$. An NFW profile is fit to the unbinned tangential shear data and shown as the black line. For illustrative purposes, the displayed profile uses a single effective photometric redshift, while the fitting procedure utilized the individual redshifts for all sources. The fit mass $M_{\mathrm{500}}$ and concentration $c$ values are listed above the right panel.}
	\label{wl_result}
\end{figure*}

\section{Results}\label{sec6}
\subsection{WL mass reconstruction}

Before incorporating an SL analysis, we performed a WL-only mass reconstruction of the cluster to probe its overall mass distribution by running \texttt{MrMARTIAN} without TNFW profiles or SL constraints. The convergence map contours are provided in the left panel of Figure \ref{wl_result}, with contours starting at $3\sigma$ and increasing in $1\sigma$ steps. This mass distribution, based solely on WL data, is highly isotropic and shows no significant secondary mass component. Additionally, the BCG is spatially consistent with the mass peak without any prior assumptions. We also estimate the cluster's total mass by fitting a spherical NFW profile to the tangential shear signal, shown in the right panel of Figure \ref{wl_result}. Given the cluster's highly isotropic mass distribution and the absence of any prominent substructure, we assume a single NFW profile centered on the BCG and fit both the mass and concentration of the cluster without imposing an $M$--$c$ relation. In this fitting, we use photometric redshifts for each background source individually rather than adopting a single effective source redshift. We define the boundary between the SL and WL regimes at the location of the Sunrise Arc and exclude any background sources within this radius. We estimate the mass of the cluster to be $M_{\mathrm{500}} = 6.7^{+1.1}_{-1.0} \times 10^{14} \, M_\odot$ with a concentration of $c = 3.8^{+1.2}_{-1.2}$. The NFW fitting utilizes shear data from both modules (A and B), whereas \texttt{MrMARTIAN} uses only data from module B, which is centered on the cluster. This choice allows for a higher-resolution convergence grid, as the resolution is limited by the background source density. Including module A does not significantly increase the source density and therefore does not justify increasing the grid resolution beyond $100 \times 100$. 

The WL mass estimate for the cluster is not consistent with the mass of $M_{\mathrm{500}} = 8.9^{+0.7}_{-0.7} \times 10^{14} \, M_\odot$ estimated in \cite{planck-2016A&A...594A..27P} using the Sunyaev-Zeldovich (SZ) effect. Even when imposing the $M$--$c$ relations from \cite{diemer2019ApJ...871..168D} and \cite{ishiyama2021MNRAS.506.4210I}, we estimate the mass to be $M_{\mathrm{500}} = 6.4^{+0.4}_{-0.4} \times 10^{14} \, M_\odot$ and $M_{\mathrm{500}} = 6.2^{+0.4}_{-0.4} \times 10^{14} \, M_\odot$, respectively. The associated concentrations are $4.2 \pm 1.0$ and $4.3 \pm 1.0$, which are both consistent with the fit concentration value shown in Figure \ref{wl_result}. Additionally, to account for the small offset between the BCG and the convergence peak, we shifted the center of the NFW profile and re-estimated the cluster mass---finding no significant change. 

The discrepancy between the SZ- and WL-derived mass estimates is likely driven by systematic biases affecting both mass estimation techniques. The SZ mass estimate is based on the Comptonization parameter $Y_{\rm SZ}$, which measures the integrated electron pressure along the line-of-sight. If a cluster is undergoing merger activity or is elongated in the line-of-sight direction, the adopted $Y_{\rm SZ}$--$M$ scaling relation can result in a mass overestimation. WHL0137-08 shows evidence of a disturbed or mixed dynamical state \citep{jimenez-Teja_2021, campitiello}, and its isotropic projected WL mass distribution suggests that any significant merger activity or elongation is likely in the line-of-sight direction. Therefore, the mass discrepancy may reflect the dynamical state of the cluster.

On the other hand, the WL mass estimate is subject to numerous systematics, including uncertainties in shape measurements, photometric redshift calibrations, and triaxiality---the last of which would be consistent with the proposed merger activity. Additionally, while the WL mass distribution is highly isotropic, the structure of the Sunrise Arc indicates a more complex substructure than what a single spherical NFW profile can adequately represent. Therefore, the disagreement between SZ- and WL-derived mass estimates---regardless of its underlying cause---is not unexpected and well within reason.

\subsection{WL+SL results}\label{wlslresults}

\begin{deluxetable*}{c|ccccc}\label{selected_models} 
\tablecaption{Selected \texttt{MrMARTIAN} models}
\tablehead {
\colhead{Label} &
\colhead{(M) Multiple-image systems\textsuperscript{a}} &
\colhead{(G) Galaxy selection\textsuperscript{b}} &
\colhead{$\Delta_{\rm rms, \, lens}$} &
\colhead{$\Delta_{\rm rms, \, source}$} &
\colhead{Earendel $|\mu|$\textsuperscript{c}}
}
\startdata
M10-G5$^{\dag}$ & 1.1(a, b); 1.1(c, d, e); 1.2; 1.7(a, c, d) & B, D, E, F, G, H, I, L, N, P & $0\farcs17$ & $0\farcs03$ & 641\\
M5-G5$^{\dag}$ & 1.1(c, d); 1.2; 1.7(a, c, d) & B, D, E, F, G, H, I, L, N, P & $0\farcs19$ & $0\farcs03$ & 269 \\
\textbf{M5-G1}$^{*}$ & 1.1(c, d); 1.2; 1.7(a, c, d) & F & $0\farcs24$ & $0\farcs06$ & 43, 67\textsuperscript{d}\\
M9-G5$^{\dag}$ & 1.1(a, b); 1.1(c, d); 1.2; 1.7(a, c, d) & B, D, E, F, G, H, I, L, N, P & $0\farcs25$ & $0\farcs03$ & 1734 \\
M1-G6$^{\dag}$ & 1.2; 1.7(a, c, d) & C, D, E, F, G, H, I, J, K, N, P & $0\farcs25$ & $0\farcs03$ & 30 \\
M6-G5$^{\dag}$ & .1(c, d, e); 1.2; 1.7(a, c, d) & B, D, E, F, G, H, I, L, N, P & $0\farcs29$ & $0\farcs03$ & 141\\
M6-G1 & 1.1(c, d, e); 1.2; 1.7(a, c, d) & F & $0\farcs28$ & $0\farcs07$ & 65 \\
M1-G1 & 1.2; 1.7(a, c, d) & F & $0\farcs30$ & $0\farcs07$ & 68\\
\enddata
\tablecomments{The set of models remaining after $\Delta_{\rm rms, \, lens} < 0\farcs3$ (before rounding) and $\Delta_{\rm rms, \, source} < 0\farcs1$ cuts, listed in order from lowest to highest $\Delta_{\rm rms, \, lens}$. The model labels indicate the chosen multiple-image systems and TNFW profiles. For example, model M1-G1 corresponds to the first set of multiple-image systems and the first TNFW profile selection listed in Table \ref{MrMARTIAN_setups}. \textsuperscript{a}Semicolons separate multiple-image systems, while commas separate images within the same system. \textsuperscript{b}All TNFW profile selections include a profile centered on the BCG. \textsuperscript{c}Fiducial model magnifications for Earendel. \textsuperscript{d}Lower magnification is the median from 100 mass reconstructions. Due to the presence of two local minima, the magnification for the best model converges to two different values. $^{\dag}$Models producing unphysical mass distributions. $^*$Our best model, chosen based on its low lens-plane scatter and physically viable mass distribution.}
\end{deluxetable*}

When incorporating SL data, we first tested only the free-form component ($100 \times 100$ grid resolution) of \texttt{MrMARTIAN} to evaluate the lens model's performance in the absence of added TNFW profiles. This free-form model alone was unable to converge multiple-image systems along the Sunrise Arc without introducing significant artifacts. We then introduced various configurations of TNFW profiles, including the original systems from W22a. Different combinations of the images from system 1.1 were used, with no assumption that Earendel lies along the critical curve. 

The impact of including TNFW profiles is evident when examining source-plane scatter. For example, model M5-G1 from Table \ref{selected_models}, which includes two TNFW profiles, achieves a source-plane scatter of $\Delta_{\rm rms, \, source} \approx 0\farcs06$, whereas removing these profiles and using only the grid component increases the scatter to $\Delta_{\rm rms, \, source} \approx 6\farcs1$. Model M9-G5 exhibits a similar trend, with $\Delta_{\rm rms, \, source} \approx 5\farcs5$ when excluding the profiles. The significant reduction in source-plane scatter when incorporating TNFW profiles highlights the effectiveness of our hybrid lens-modeling approach in combining WL and SL datasets from the WHL0137-08 field.

In total, we tested 19 different multiple-image system assignments and 14 different cluster member galaxy selections, resulting in 133 total model setups. This number is lower than the simple product ($19 \times 14 = 266$) of these values since TNFW profiles at the positions of galaxies Q and R (see Figure \ref{halos}) can only be used when multiple-image system 2 is considered. The full set of models is outlined in Table \ref{MrMARTIAN_setups}, with relevant galaxy positions labeled in Figure \ref{halos}.

To narrow down the sample of models, we selected those that produce lens-plane scatter of $\Delta_{\rm rms, \, lens} < 0\farcs3$ (15 pixels) and source-plane scatter of $\Delta_{\rm rms, \, source} < 0\farcs1$ (5 pixels). These thresholds were chosen based on two considerations. First, because we are investigating the magnification of Earendel---which is highly sensitive to the lens model---the lens-plane scatter limit should be lower than typical values reported in the literature ($0\farcs31$--$1\arcsec$; \citealt{chaHFF2023ApJ...951..140C}). Second, source-plane scatter can be artificially minimized in the model by increasing the magnification, so a low---but not overly constraining---threshold is sufficient. We find that our models reconstructed using the multiple-image system assignment from W22a achieve a typical lens-plane scatter\footnote{W22a and W22b did not quantify lens-plane scatter.} of $\Delta_{\rm rms, \, lens} \approx 1''$, with none of these models meeting our selection criteria. Additionally, any models that incorporate multiple-image system 2 exhibit unphysical mass distributions and similarly large lens-plane scatter. This is in contrast with W22a, whose lens models relied on system 2 to reproduce the full length of the Sunrise Arc. 

All eight models remaining after the cut, shown in Table \ref{selected_models}, utilize our newly identified multiple-image system $1.2$, further supporting its validity as a true system. Of these final models, five$^\dag$ exhibit unphysical mass distributions due to improper TNFW profile parameters. For instance, model M9-G5, which produces the highest magnification estimate at Earendel's position, has its highest convergence peak centered on galaxy N (labeled in figure \ref{halos}) rather than the BCG. Moreover, the TNFW profile assigned to this galaxy has a concentration of $c\approx14.3$, scale radius of $r_{\rm s} \approx 0.04$ Mpc, and a truncation ratio (i.e., $r_{\rm truncation}/r_{\rm s}$) of $\tau \approx 0.13$. These parameters yield an unrealistically compact TNFW profile that is inconsistent with the observed properties of the associated galaxy. The corresponding convergence map and critical curves for model M9-G5 are shown in Figure \ref{m9-g5}.

\begin{figure}[tp]
    \centering
    \includegraphics[width=\columnwidth]{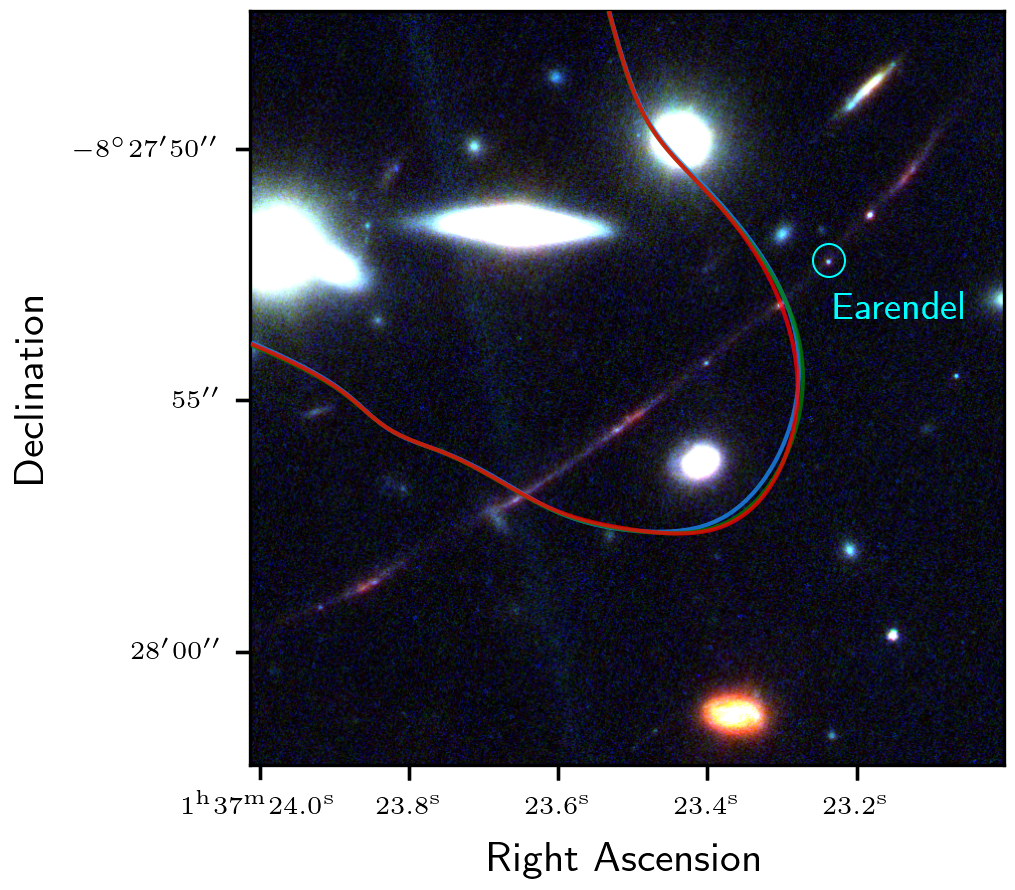}
    \caption{Critical curves at $z=6.2$ produced by models M5-G1 (red), M6-G1 (green), and M1-G1 (blue). The westernmost crossing location of the critical curve along the arc is shifted significantly away from Earendel ($1\farcs15$--$1\farcs30$) when compared to models from the literature.}
    \label{final_critcurves}
\end{figure}

After discarding unphysical models, we are left with three viable models (M5-G1, M6-G1, and M1-G1), each utilizing two TNFW profiles: one centered on the BCG and the other on galaxy F. This exclusion also narrows the magnification range from 43--1734 to 43--68, which is two to three orders of magnitude lower than the estimates reported in W22a and W22b. This reduction results from a significant shift in the critical curve crossing location, as shown in Figure \ref{final_critcurves}. W22a reports a distance between Earendel and the critical curve of $0\farcs02$--$0\farcs036$, and as small as $0\farcs005$ in W22b, depending on the lens model. In contrast, our best models place this separation between $1\farcs15$ and $1\farcs3$. It is important to note that our distance estimates are derived differently: W22a and W22b infer the separation as part of their magnification estimation procedure, whereas we measure it directly from high-resolution magnification maps. This approach is enabled by the significant offset between Earendel and the critical curve in our three viable models.

Interestingly, these three models exhibit critical curves that pass through image 1.1b rather than Earendel, which would suggest that this image may be a candidate star rather than Earendel. However, although the majority of our models predict a low magnification for Earendel, the critical curve crossing location along the Sunrise Arc is not sufficiently well constrained to justify this assertion.

 We select model M5-G1$^{*}$, listed in Table \ref{selected_models}, as our best model as it exhibits both low lens-plane scatter and a physically viable mass distribution. We make this lens model publicly available, including the magnification map, convergence map, and deflection maps, all scaled to Earendel's redshift of $z=6.2$\footnote{\url{https://doi.org/10.5281/zenodo.15110933}}.

\section{Discussion} \label{sec7}

\subsection{Lens models}

\begin{figure*}[tp]
\centering
	\includegraphics[width=\textwidth]{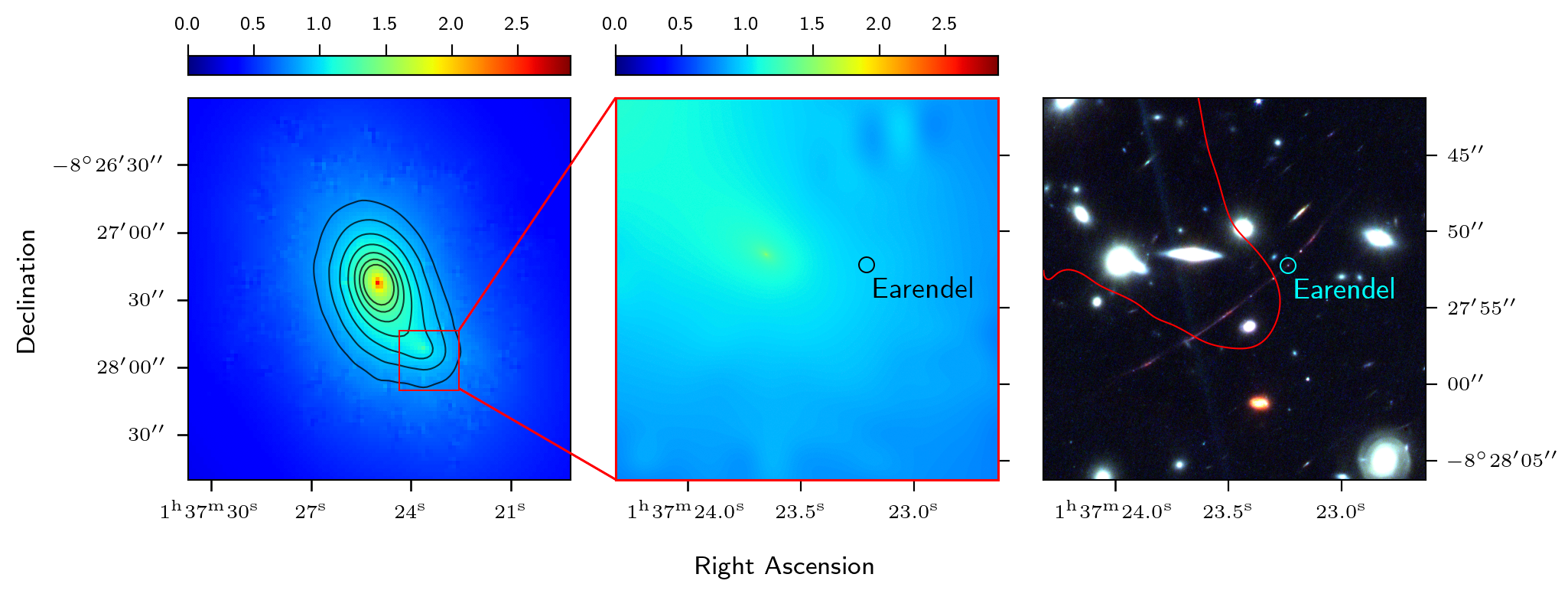}
	\caption{\texttt{MrMARTIAN} modeling results for our best model M5-G1, which shows smooth substructure nearby the Sunrise Arc. Left: $100 \times 100$ \texttt{MrMARTIAN} convergence map and contours, with the region surrounding the Sunrise Arc outlined in red. The contours correspond to this same convergence map smoothed with a Gaussian kernel where $\sigma \approx 2\farcs5$. Middle: High-resolution convergence map for the arc region, outlined with a red box in the left panel. Note that the high-resolution TNFW profiles are based on their analytic form and not a result of interpolation. Right: Critical curve at redshift $z=6.2$ in the arc region.}
	\label{m5-g1}
\end{figure*}

The TNFW profile centered on galaxy F's position in our best model, M5-G1 (as well in models M6-G1 and M1-G1), exhibits a low concentration ($\mytilde$0.8) and large scale radii ($\mytilde$1 Mpc), resulting in a smooth mass distribution as shown in Figure \ref{m5-g1}. This TNFW profile likely represents the larger-scale substructure in the cluster rather than a single galaxy mass profile---a hypothesis supported by the concentration of cluster galaxies in this diffuse TNFW halo. Interestingly, the TNFW profile centered on the BCG converges to an elliptical configuration across all models that utilize TNFW profiles, including our best model, suggesting that an elliptical halo provides a better fit to the SL constraints than a circular one.

A consequence of using a single TNFW profile centered at galaxy F's position is that the critical curve's path is not significantly affected by mass associated with other galaxy positions. As shown in Figure \ref{final_critcurves}, all three of the shown viable models produce critical curves that are tangent to galaxy G. Given that this galaxy is the closest bright cluster member galaxy nearby Earendel---and could potentially shift the critical curve if explicitly modeled---we created additional models with TNFW profiles at the position of this galaxy and that of nearby bright galaxies D and F. We used the same multiple image systems as our viable models shown in Figure \ref{final_critcurves} (M5, M6, M1), as well as the systems from W22a (M3, excluding system 2), resulting in four additional models. In all four cases, the TNFW profile at galaxy G's position was effectively removed during the minimization, with its concentration pushed to the lower boundary ($c=0.01$). The exclusion of this TNFW mass profile in \texttt{MrMARTIAN} modeling does not necessarily imply that galaxy G does not influence the magnification of Earendel, particularly because the grid component of the algorithm can compensate for its removal. However, it does suggest that further testing is required to better constrain the mass distribution in this region.

The simplest model is M1-G1, only utilizing multiple-image systems 1.7(a, c, d) and 1.2. Models M5-G1 and M6-G1 include images from system 1.1, with M6-G1 treating Earendel (1.1e) as part of a triply-imaged system. Before beginning the modeling process, we anticipated that images 1.1c and 1.1d may be part of the same system given their similar color and their observed parity along the arc. This parity also suggests that one of 1.1a, 1.1b, or Earendel is an image in the same system, with our models pointing to Earendel being the most likely candidate. As mentioned in Section \ref{mult_syst}, Earendel also shares a similar color to images 1.1c and 1.1d, which can be seen in Figure \ref{sunrise_systems}.

To provide more robust estimates for Earendel's magnification, we performed 100 additional mass reconstructions using the multiple-image system and galaxy position selections from our best model, M5-G1. For each reconstruction, we perturbed the initial TNFW profile parameters and initialized the convergence map grid with values drawn from a uniform distribution centered at $\kappa=0$ and spanning $-0.1 < \kappa < 0.1$. From the resulting mass maps, we estimated Earendel's magnification to be $\mu = 42.57^{+0.35}_{-0.48}$. The consistent separation between the critical curve and Earendel across these reconstructions leads to the low uncertainty in this estimate, as the magnification gradient is shallower farther from the curve.  Our final prediction for Earendel's magnification is in the range $\mu = 43$--$67$, where 43 is the rounded median of the ensemble of models and 67 corresponds to the fiducial model value quoted in Table \ref{selected_models}. The systematic bias between the fiducial model's prediction and the ensemble median prediction results from the perturbations to the initial convergence grid, as forcing the TNFW profile parameters to match the fiducial model in each iteration does not resolve the offset in magnification. 

Two of the models from Table \ref{selected_models}$^\dag$, models M9-G5 and M10-G5, produce unphysical mass distributions and predict considerably higher magnifications at Earendel's position, as mentioned before. These models treat images 1.1a and 1.1b as multiple images, and given that Earendel is located at the midpoint between them along the arc, using these images as constraints strongly favors the presence of a critical curve at Earendel's position. However, incorporating 1.1a and 1.1b as constraints consistently introduces artifacts in our mass reconstructions and increases the lens-plane scatter across all multiple-image systems. Models M9-G5 and M10-G5 also include a TNFW profile located at the position of galaxy H---the nearest galaxy to Earendel. In these models, the \texttt{MrMARTIAN} minimization assigns this TNFW profile an unrealistically high mass given its luminosity, significantly altering the path of the critical curve. An example of this can be seen in the middle and right panels of Figure \ref{m9-g5}. While this faint galaxy may influence lensing features along the arc, its mass profile should not be comparable to those of the surrounding bright cluster member galaxies. Since \texttt{MrMARTIAN} uses flat priors for TNFW profile parameters that are not informed by photometry, this is sometimes unavoidable.

To demonstrate that the low magnification predicted for Earendel in our best model is not an outlier, we provide the full set of magnification estimates from all 133 of our models in Figure \ref{mag_hist}. The median and $1\sigma$ range for Earendel's magnification across all models is $112^{+269}_{-61}$, which is consistent with our best model's fiducial estimate of $\mu = 67$. However, many of these lens models exhibit unphysical mass distributions, making their magnification estimates less reliable. Some models should be penalized or down-weighted based on a combination of their complexity and ability to reproduce multiple-image positions (e.g., using the Bayesian Information Criterion), which is not done here. If such weighting were taken into account, the upper $1\sigma$ bound on Earendel's magnification could be lower than the current estimate.

We recognize that our magnification estimates differ from those of W22a, even under nominally similar conditions (models M12-G5, M3-G5). However, several key methodological differences naturally contribute to this discrepancy:
\begin{itemize}
    \item Given the truncation of TNFW profiles and inclusion of a free-form component, we restrict our parametric modeling to the BCG and galaxies located near the Sunrise Arc.
    \item The hybrid algorithm used in W22a, \texttt{WSLAP+}, explicitly applies mass-to-light constraints, whereas our approach employs uniform priors for TNFW profile parameters.
    \item The inclusion of WL constraints affects the global mass distribution, indirectly influencing local lensing properties in Earendel's vicinity. 
\end{itemize}
These methodological choices, while differing from previous SL-only models, provide independent magnification constraints with transparent assumptions and quantified uncertainties.

\subsection{Earendel's stellar candidacy}

\begin{figure}[tp]
    \centering
    \includegraphics[width=\columnwidth]{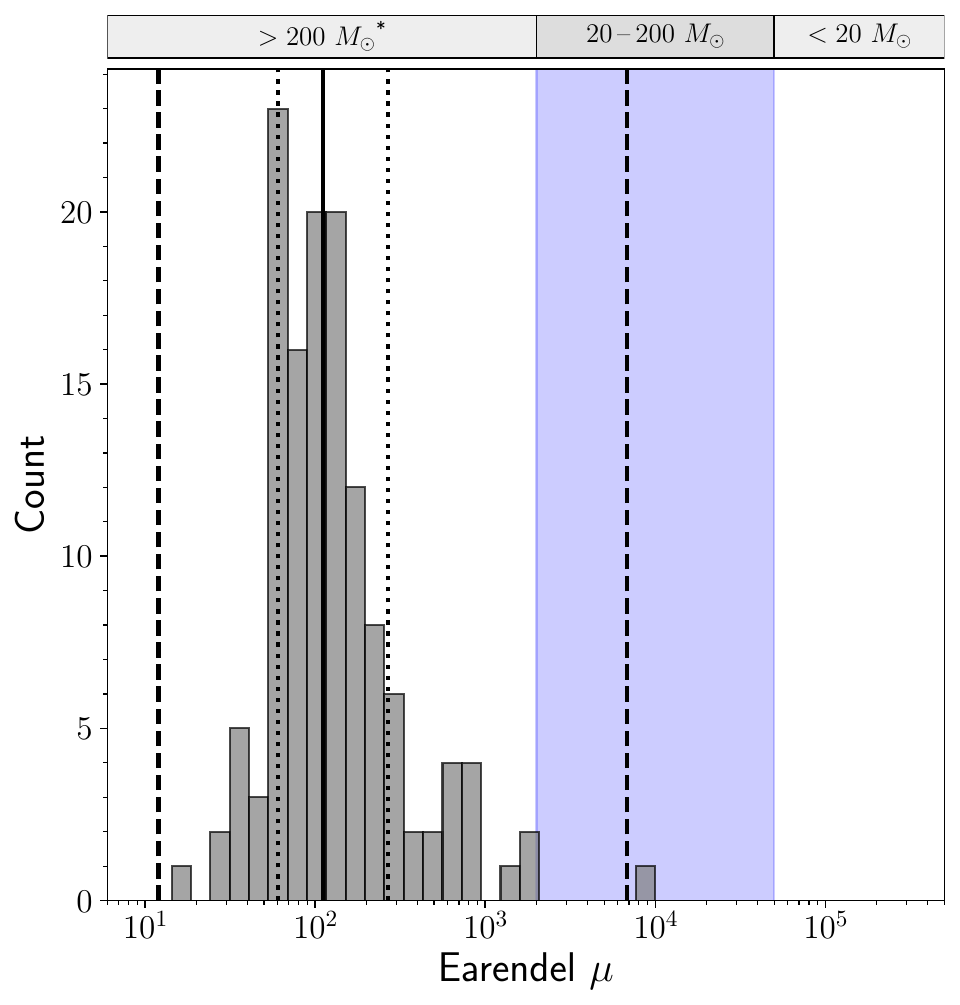}
    \caption{Magnification distribution at the location of Earendel from our 133 lens models. The blue-shaded region marks the plausible magnification range reported in W22b ($2\mu = 4{,}000$--$100{,}000$), where the upper limit reflects the highest achievable magnification indicated by the microlensing analysis in W22a. The stellar mass intervals at the top of the figure are based on Figure 5 of W22b, which maps this magnification range onto an H-R diagram to infer zero-age main sequence (ZAMS) mass estimates. $^*$The mass in this bin can significantly exceed $200~M_{\odot}$; for example, stars with $\log(L) \approx7$--$7.5~L_{\odot}$ have a mass $M\gtrsim500~M_{\odot}$ in the H-R diagram from W22b.}
    \label{mag_hist}
\end{figure}

W22b used their magnification estimates to argue that a single star could reproduce Earendel's observed properties with a ZAMS mass between $20$--$200~M_{\odot}$, corresponding to their full range of plausible magnifications. This range is illustrated as the blue-shaded region in Figure \ref{mag_hist}. In contrast, we find that only a single lens model gives Earendel a magnification within this range, with this value being a 3$\sigma$ outlier. Moreover, the associated lens model exhibits a highly unphysical mass distribution---exhibiting overly compact TNFW profiles with excessively high convergence peaks.

W22b also point out that a magnification of $2\mu = 7{,}000$ is 25 times more probable than $2\mu = 35{,}000$, since the probability of an object having a given magnification scales as $P(>\!\mu) \propto \mu^{-2}$. They further highlight challenges associated with extremely high-mass, high-luminosity stars: for instance, a $\mytilde15{,000}~\rm K$ star with a luminosity above $~10^6~L_{\odot}$ would exceed the Humphreys--Davidson limit, an empirical ceiling above which no stars have been observed in the local universe. Using a simplified approach, we estimate Earendel's intrinsic luminosity to be $\log(L) \approx 8.6$--$8.9~L_{\odot}$, well beyond this limit. This estimate is based on the F410M filter flux reported in Table 1 of W22b, our magnification range of $\mu = 43$--$67$, and a bolometric correction of $-1.5$---chosen to approximately reproduce the luminosity ranges in W22b under their higher magnification assumptions. Adopting a standard $M$--$L$ relation for high-mass stars of $\mytilde2$, the luminosity estimates correspond to a mass range of roughly $10^{4.3}$–$10^{4.5}~M_{\odot}$.

We emphasize that this is an order-of-magnitude estimate, and while our simplified luminosity estimate was calibrated to be consistent with the results from W22b, their analysis incorporates detailed stellar atmosphere models and photometric SED fitting. Our calculation is intended to illustrate that the interpretation of Earendel as a single star---or even a small group of stars---is improbable given the magnifications from our best-performing lens models, as this would imply an implausibly high luminosity and stellar mass.

\section{Conclusion} \label{sec8}

Using JWST imaging data, we find that the multiple-image systems within the Sunrise Arc and near Earendel likely differ from those assigned in previous studies. We have also conducted a combined WL and SL analysis of the galaxy cluster WHL0137-8 using the hybrid lens modeling algorithm \texttt{MrMARTIAN}. In constructing the WL dataset, we carefully reduced the F200W filter data, modeled the observed PSF, and applied an appropriate shear bias correction to ensure an accurate result. To combine the WL and SL constraints and perform a unified lensing analysis, we included SL multiple-image positions as constraints and introduced TNFW profiles into the reconstruction field. In doing so, we achieved multiple-image source- and lens-plane scatters within $0\farcs1$ and $0\farcs3$, respectively.

Though some of our models predict high magnification at Earendel's position, the corresponding mass distributions are highly unphysical. In our best model, the critical curve does not intersect or pass near Earendel, whose magnification is insufficient to support its identification as a single star at $z\approx6.2$ with a plausible stellar mass. It is more probable that Earendel is a compact cluster of stars or a globular cluster progenitor, which would require less significant magnification to account for its observed brightness.

Even without considering lens modeling, the JWST imaging data alone strongly suggests that the multiple-image system assignments in previous studies may be incomplete or require revision. Given the limited SL constraints available for this cluster, any revision to the multiple-image system assignment can significantly affect Earendel's magnification. With our revised multiple-image catalog, we encourage further investigation using these candidates with alternative lens-modeling algorithms.

\section*{Acknowledgments}
We are grateful to Kim HyeongHan for providing independent algorithms that aided in the verification of our mass estimates and to Kyle Finner and Wonki Lee for insightful discussions regarding the WL analysis results. We also extend our thanks to the Cosmic Spring team for making their photometric redshift catalog for this field public. This work is based on observations made with the NASA/ESA/CSA JWST and downloaded from the Mikulski Archive for Space Telescopes (MAST) at the Space Telescope Science Institute (STScI). The data described here may be obtained from the MAST archive at \dataset[doi:10.17909/q1mq-d204]{https://dx.doi.org/10.17909/q1mq-d204}. MJJ acknowledges support for the current research from the National Research Foundation (NRF) of Korea under the programs 2022R1A2C1003130 and RS-2023-00219959.

\vspace{5mm}
\facilities{JWST(NIRCam)}

\software{astropy \citep{2013A&A...558A..33A,2018AJ....156..123A},  
          Source-Extractor \citep{bertin-1996A&AS..117..393B},
          JWST Calibration Pipeline \citep{bushouse_2024_12692459},
          STPSF \citep{webbpsf},
          Colossus \citep{colossus2018ApJS..239...35D},
          scipy \citep{scipy2020NatMe..17..261V}
          }

\appendix

\twocolumngrid
\counterwithin{figure}{section}

\section{Data reduction}
\label{app_data_reduction}
The \texttt{young-jwstpipe} JWST calibration pipeline is intended to simplify and expedite the creation of clean JWST NIRCam mosaic images using the Python interface of the standard JWST calibration pipeline \citep{bushouse_2024_12692459}. It allows for the combination of multiple observations into one mosaic image by generating custom associations, applying additional calibration steps beyond those in the default pipeline, and aligning images across different filters using an astrometric reference catalog derived from a filter chosen by the user. 

The first additional step introduced is wisp removal, which is implemented by scaling and subtracting wisp templates from NIRCam exposures. For this work, we utilized the initial version of the algorithm created by Christopher Willmer and the templates provided in the JWST User Documentation \citep{2016jdox.rept......}, with minor modifications to integrate the code into our pipeline. The \texttt{young-jwstpipe} pipeline has since been updated to utilize the most recent version of the wisp removal algorithm, created by Ben Sunnquist, and templates collected from approximately two years of flight data.

The next step involves a modified version of Micaela Bagley’s \texttt{remstriping} algorithm. We adapted this algorithm to improve performance in cluster fields, which often contain extended, bright sources, whereas the original \texttt{remstriping} method was designed for blank fields. Our modification makes the 1/\emph{f} noise subtraction more conservative near bright sources. Finally, we include Henry C. Ferguson’s \texttt{background subtraction} code to ensure a flat sky level in the final mosaic images. The original \texttt{remstriping} and background subtraction routines are accessed through the CEERS GitHub repository \citep{ceers}. More recently, an alternative 1/\emph{f} noise correction algorithm has been added to the pipeline to better preserve intra-cluster light through the use of a gradient-based technique.

\section{Sunrise Arc Denoising}\label{denoising_desc}
When denoising the Sunrise Arc, we utilized the Restormer \citep{zamir2022restormer} deep learning architecture---a vision transformer architecture that both reduces computational costs and can be applied to larger images. 

We first cropped a 950 $\times$ 800 pixel region surrounding the Sunrise Arc from the full WHL0137-08 field. This is done to prevent memory overconsumption during the denoising process and to mitigate the influence of any large-scale correlations, such as residual background gradients. The image was then segmented into $64 \times 64$ patches for which the model can utilize self-attention, which calculates the importance of each pixel relative to all other pixels in the same patch. This helps capture pixel correlations independent of distance.

To simulate realistic JWST noise in training data, we adopted the correlated noise modeling procedure from \cite{springer2020MNRAS.491.5301S}. We first extracted $37 \times 37$ pixel background patches from the image to construct the following empirical covariance matrix:
\begin{equation}
    \left(\Sigma_{\mathrm{BG}}\right)_{j,k} = \left(\frac{1}{m-1}\sum_{i=1}^{m}p_i p_i^T\right)_{j,k} \, .
\end{equation}
Here, $p_i$ denotes the vectorized background patches, and each matrix element captures the covariance between the $j^{\mathrm{th}}$ and $k^{\mathrm{th}}$ pixel in these vectorized patches. The matrix square root $L$ of $\Sigma_{\mathrm{BG}}$ satisfies $\Sigma_{BG} = L L^T$, which allowed us to map uncorrelated i.i.d. standard noise vectors into correlated vectors that match the empirical covariance:
\begin{align}
    \mathrm{Cov}(L\bm{z}) = L\mathrm{Cov}(\bm{z})L^T & = L\, I_n\, L^T \nonumber \\ & = L \, L^T = \Sigma_{\mathrm{BG}} \, .
\end{align}
The central row of $L$ was then reshaped into a convolution kernel and convolved with a standard normal noise map to create a noise distribution exhibiting JWST correlated noise features. Matrix multiplication can also be used in place of this convolution with the full matrix $L$, however, the difference between these two approaches is negligible. 

In \cite{springer2020MNRAS.491.5301S}, $7\times7$ pixel background patches were used instead of $37\times37$. We increased the patch size to capture large-scale variations in the background noise, which reduces the residual background gradient in the denoised image and prevents the mischaracterization of noise as faint sources. This value was chosen empirically, as increasing the patch size beyond $37\times37$ resulted in too few background patches to reliably estimate the noise covariance matrix. 
\begin{figure*}[b]
\centering
	\includegraphics[width=\textwidth]{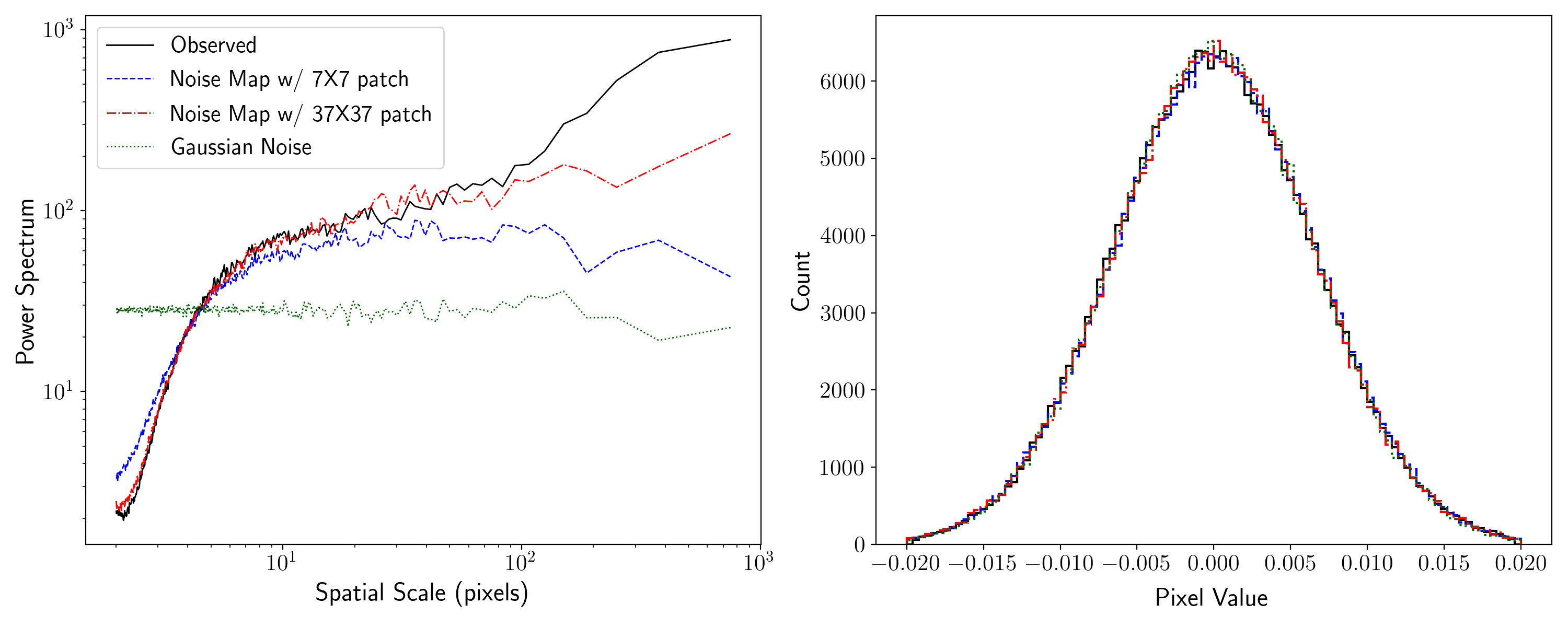}
	\caption{Left: Power spectrum of the F444W filter observed noise(black solid), the F444W correlated noise map created with 7$\times$7(blue dashed), the F444W correlated noise map created with  37$\times$37 (red dash-dot) background patches, and standard normal noise (green dotted) with the same mean and standard deviation as the observed noise. The larger background patch size results in the correlated noise power spectrum being a closer match to the observed noise power spectrum. Right: Histogram of background pixel values from noise maps and the observation. Regardless of the background patch size, the modeled correlated noise distribution follows the observed noise distribution well.}
	\label{psd_comparison}
\end{figure*}
\begin{figure*}[bp]
\centering
	\includegraphics[width=\textwidth]{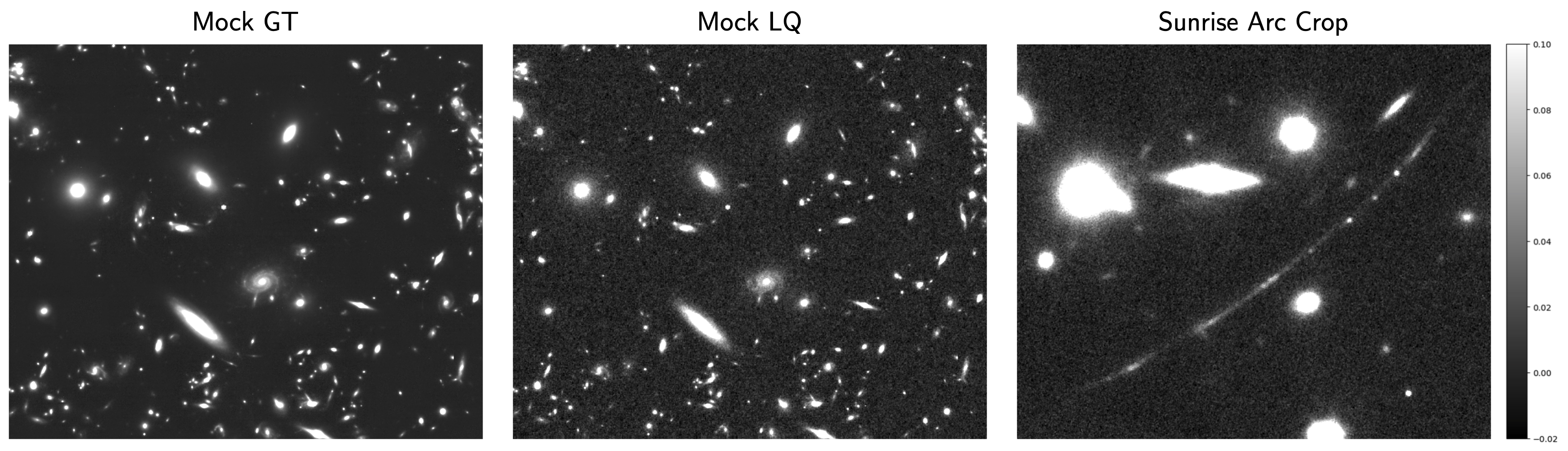}
	\caption{Mock GT image (left), mock LQ image (middle), and the cropped region around the Sunrise Arc (right) are displayed with the same scaling to benefit the comparison. }
	\label{mock_image}
\end{figure*}
The left panel of Figure \ref{psd_comparison} compares the power spectra of the observed F444W noise and modeled correlated noise (from both $7\times7$ and $37\times37$ patches), showing improved spectral agreement with the larger patch size. The right panel shows that both patch sizes reproduce the observed pixel value distribution well.

To create low-quality (LQ) images for training, we added galaxies and stars sampled from other JWST observations at similar wavelengths to the correlated noise maps, taking care to ensure that the added source cutouts did not contaminate the noise model. Cutout sizes for each source were defined as $12a + 10$ pixels, where $a$ is the semi-major axis length produced by Source-Extractor. After masking nearby sources, each cutout was background-subtracted and rescaled so that its RMS was one-tenth of the modeled noise RMS. To compensate for this dimming, we selected sources initially brighter than the objects in the WHL0137-08 field.

The ground-truth (GT) images were created by adding the same sources to a zero background, with Gaussian noise added to empty regions based on sampled source noise statistics. Figure~\ref{mock_image} shows examples of GT, LQ, and Sunrise Arc images. The mock images contain more bright galaxies due to our selection process, which ultimately benefits training: since noise dominates many pixels in the Sunrise Arc region, the inclusion of more galaxies helps the model learn intrinsic features and smooth transitions to the background.

From 100 GT-LQ mock image pairs, we extracted 100,000 random $64\times64$ postage stamp pairs. The dataset was split into 80,000 training, 10,000 validation, and 10,000 test image pairs. We trained the Restormer model for 60,000 iterations (batch size of 32) on an RTX 3090 GPU, with training completed in approximately 9 hours. During inference, the Sunrise Arc image was segmented into overlapping 64$\times$64 postage stamps using a 16-pixel step size. The final denoised image was generated by taking the median at each pixel across overlapping outputs.

Denoising and training were performed independently for each filter to account for differences in noise characteristics, PSF, and depth.  For each filter group (F090W/F115W, F150W/F200W, F277W/F356W, F410M/F444W), sources were drawn from the F090W, F200W, F277W, and F444W filters, respectively. All training images were coadded using the \texttt{young-jwstpipe} pipeline with a square kernel, \texttt{pixfrac} = 0.75, and an output pixel scale of $0\farcs02 ~\rm pixel^{-1}$ to match the process used for the full WHL0137-08 mosaic.

\section{Extended Data}

\begin{figure*}[tp]
\centering
	\includegraphics[width=\textwidth]{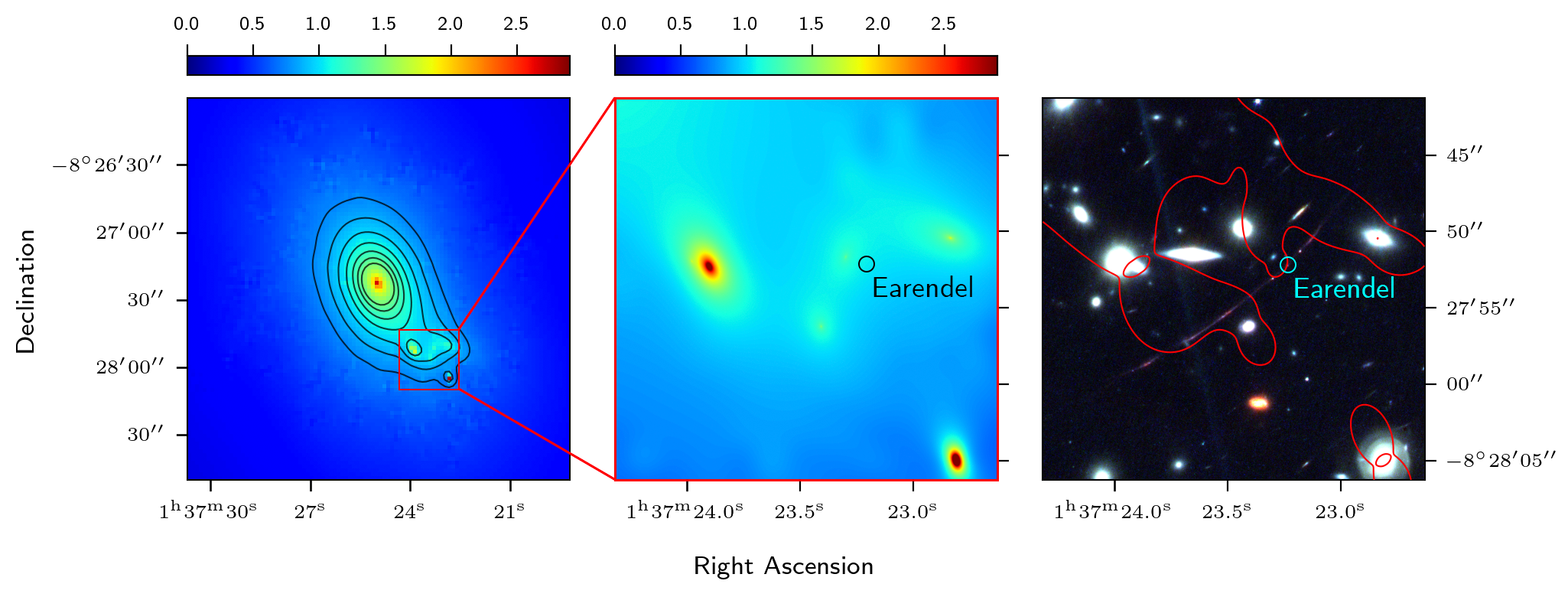}
	\caption{Example of an unphysical \texttt{MrMARTIAN} model mass distribution, with this being model M9-G5. Left: $100 \times 100$ \texttt{MrMARTIAN} convergence map and contours, with the region surrounding the Sunrise Arc outlined with a red box. The contours correspond to this same convergence map smoothed with a Gaussian kernel where $\sigma \approx 2\farcs5$. Middle: High-resolution convergence map for the arc region, outlined with a red box in the left panel. Note that the high-resolution TNFW profiles are based on their analytic form and not a result of upsampling. Right: Critical curves at redshift $z=6.2$ in the arc region.}
	\label{m9-g5}
\end{figure*}

\begin{deluxetable*}{c|cc|c}\label{MrMARTIAN_setups} 
\centering
\tablecaption{\texttt{MrMARTIAN} model setups}
\tablehead {
\colhead{Label} &
\colhead{(M) Multiple-image systems\textsuperscript{a}} &
\colhead{Label} &
\colhead{(G) Galaxy selection\textsuperscript{b}}
}
\startdata
M1 & 1.2; 1.7(a, c, d) & G1 & F\textsuperscript{c} \\
M2 & 1.2; 1.7(a, b, c, d) & G2 & F, H, L\textsuperscript{d} \\
M3 & 1.1(a, b, c); 1.7(a, b, c) & G3 & F, I, N, P\textsuperscript{e} \\
M4 & 1.1(a, b); 1.2; 1.7(a, c, d) & G4 & B, D, E, F, G, H, I, N, P\textsuperscript{f} \\
M5 & 1.1(c, d); 1.2; 1.7(a, c, d) & G5 & B, D, E, F, G, H, I, L, N, P\textsuperscript{g} \\
M6 & 1.1(c, d, e); 1.2; 1.7(a, c, d) & G6 & C, D, E, F, G, H, I, J, K, N, P\textsuperscript{h} \\
M7 & 1.1(a, b, c); 1.2; 1.7(a, c, d) & G7 & A, B, C, D, E, F, G, H, I, J, K, M, N, O, P\textsuperscript{i} \\
M8 & 1.1(c, b, e); 1.2; 1.7(a, c, d) & G8 & F $+$ Q, R\textsuperscript{j} \\
M9 & 1.1(a, b); 1.1(c, d); 1.2; 1.7(a, c, d) & G9 & F, H, L $+$ Q, R\\
M10 & 1.1(a, b); 1.1(c, d, e); 1.2; 1.7(a, c, d) & G10 & F, I, N, P $+$ Q, R\\
M11 & 1.1(a, b, c); 1.7(a, b, c); 2(a, b) & G11 & B, D, E, F, G, H, I, N, P $+$ Q, R\\
M12 & 1.1(a, b, c); 1.7(a, b, c); 2(a, b, c)$^{\dag}$ & G12 & B, D, E, F, G, H, I, L, N, P $+$ Q, R\\
M13 & 1.2; 1.7(a, c, d); 2(a, b) & G13 & C, D, E, F, G, H, I, J, K, N, P $+$ Q, R\\
M14 & 1.2; 1.7(a, c, d); 2(a, b, c) & G14 & A, B, C, D, E, F, G, H, I, J, K, M, N, O, P $+$ Q, R\\
M15 & 1.1(c, d); 1.2; 1.7(a, c, d); 2(a, b, c) & & \\
M16 & 1.1(a, b); 1.2; 1.7(a, c, d); 2(a, b, c) & & \\
M17 &  1.1(a, b, c); 1.2; 1.7(a, c, d); 2(a, b, c) & & \\
M18 & 1.1(c, d, e); 1.2; 1.7(a, c, d); 2(a, b, c) & & \\
M19 & 1.1(a, b); 1.1(c, d, e); 1.2; 1.7(a, c, d); 2(a, b, c) & & \\
\enddata

\tablecomments{The full set of models consists of all possible combinations of the relevant multiple-image system configurations and TNFW profiles placed at the positions of selected galaxies. \textsuperscript{a}Semicolons separate multiple-image systems, while commas separate images within the same system. \textsuperscript{b}All TNFW profile selections include a profile centered on the BCG. \textsuperscript{c}Profile F is selected to represent the characteristic position of bright cluster members near the Sunrise Arc. \textsuperscript{d}Profiles H and L are added given their proximity to the arc. \textsuperscript{e}This selection is intended to bracket the arc in all cardinal directions. \textsuperscript{f}Cluster members near the arc are selected based on an analysis of lens models from W22a. \textsuperscript{g}Cluster members near the arc, including galaxy L, selected based on lens models from W22a. \textsuperscript{h}The 11 most prominent cluster members near the arc, selected based on photometric redshifts. \textsuperscript{i}All 15 cluster member galaxies within 50kpc of the Sunrise arc, selected based on photometric redshifts. \textsuperscript{j}TNFW profiles placed at the positions of galaxies Q and R are only used in lens models using multiple-image system $2$. $^\dag$The multiple-image system from W22a, identical to M3 except with system 2 added. M11 is the same as M12 except it excludes image 2c, which was a test performed in an attempt to converge system 2.}
\end{deluxetable*}

\begin{deluxetable}{c|cc}\label{multiple_images} 
\centering
\tablecaption{Multiple images}
\tablehead {
\colhead{Image} &
\colhead{Right Ascension} &
\colhead{Declination}
}
\startdata
1.1a & 24.346591 & -8.464253 \\
1.1b & 24.347093 & -8.464749 \\
1.1c & 24.349662 & -8.466417 \\
1.1d & 24.347503 & -8.465067 \\
1.1e & 24.346822 & -8.464511 \\
\hline
1.2a & 24.346396 & -8.464045 \\
1.2b & 24.347901 & -8.465367 \\
1.2c & 24.349429 & -8.466311 \\
\hline
1.7a & 24.346351 & -8.463995 \\
1.7b & 24.348570 & -8.465825 \\
1.7c & 24.349359 & -8.466278 \\
1.7d & 24.348005 & -8.465436 \\
\hline
2a & 24.345958 & -8.457367 \\
2b & 24.346171 & -8.458089 \\
2c & 24.346522 & -8.459075 \\
\enddata
\tablecomments{Multiple images used in this work.}
\end{deluxetable}

\begin{figure*}[t]
    \centering
    \includegraphics[width=0.7\textwidth]{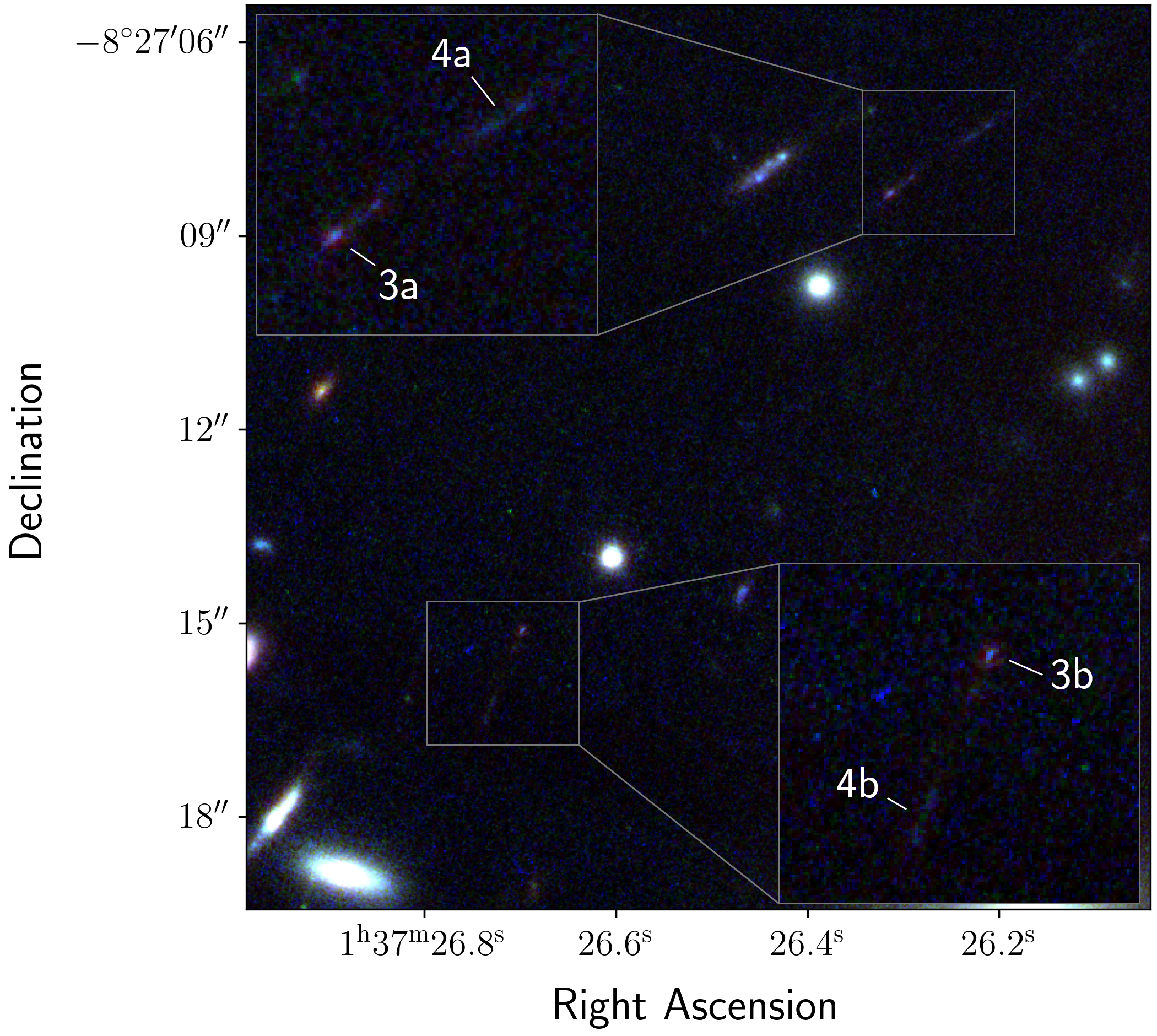}
    \caption{Region R1 from Figure \ref{full_field}, showing newly identified multiple-image candidates (3a, 3b; 4a, 4b) at $z \approx 6.6$ in the WHL0137-08 field.}
    \label{extra_system}
\end{figure*}
    
\clearpage
\bibliographystyle{apj}
\bibliography{main}

\end{document}